\documentclass[aps,prx,preprintnumbers,twocolumn,nofootinbib,longbibliography,notitlepage]{revtex4-1}
\usepackage{amssymb,amsmath,amsthm}
\usepackage[dvips]{graphicx}
\usepackage{verbatim}
\usepackage{epsfig}
\usepackage{color}
\usepackage{comment}
\usepackage{subfigure}

\newcommand\kT{k_{\mathrm{B}}T}

\usepackage{graphicx}
\usepackage{bm}  
\usepackage{epsfig}
\usepackage{mathtools}
\usepackage{verbatim} 

\begin{document}

\title{Density--nematic coupling in isotropic linear polymers}

\date{\today}

\author{Aleksandar Popadi\' c$^{1}$, Daniel Sven\v sek$^{2,*}$, Rudolf Podgornik$^{2,3,4,5}$, and Matej Praprotnik$^{1}$}

\affiliation{
\mbox{$^1$Laboratory for Molecular Modeling, National Institute of Chemistry, SI-1001 Ljubljana, Slovenia}\\
\mbox{$^2$Department of Physics, Faculty of Mathematics and Physics, University of Ljubljana, SI-1000 Ljubljana, Slovenia}\\
\mbox{$^3$Department of Theoretical Physics, J.\ Stefan Institute, SI-1000 Ljubljana, Slovenia}\\
\mbox{$^4$School of Physical Sciences and Kavli Institute for Theoretical Sciences, University of Chinese Academy of Sciences,}\\ \mbox{Beijing 100049, China}\\ 
\mbox{$^5$CAS Key Laboratory of Soft Matter Physics, Institute of Physics, Chinese Academy of Sciences, Beijing 100190, China}\\
$^{*}$email: daniel.svensek@fmf.uni-lj.si}

\begin{abstract}

\noindent 
Linear polymers and other connected ``line liquids'' exhibit a coupling between density and equilibrium 
nematic order on the macroscopic level that gives rise to a Meyer-de Gennes {\em vectorial conservation law}. Nevertheless, isotropic linear polymer melts/solutions exhibit fluctuations of the density and of the nematic order that are not coupled by this vectorial constraint, just like for isotropic liquids composed of disconnected non-spherical particles. It takes the proper tensorial description of the nematic order in linear polymer liquids, leading to a {\em tensorial conservation law} connecting density and orientational order, that finally implicates coupled density and nematic order fluctuations, already in the isotropic system and not subject to the existence of an orientational phase transition. This coupling implies that a spatial variation of density or a local concentration gradient will induce nematic order and thereby an acoustic or osmotic optical birefringence even in an otherwise isotropic polymer melt/solution. 
We validate the theoretical conceptions by performing detailed Monte Carlo simulations of isotropic melts of ‘‘soft’’ worm-like chains with variable length and flexibility, and comparing the numerically determined orientation correlation functions with predictions of the macroscopic theory.
The methodology drawn sets forth a means of determining the macroscopic parameters by microscopic simulations to yield realistic continuum models of specific polymeric materials.
\end{abstract}
\maketitle

\section{Introduction}

\noindent
Isotropic liquids possess no macroscopic preferred direction. Nevertheless, if they consist of non-spherical microscopic elementary units like prolate, oblate or more complicated shapes, they do exhibit collective fluctuations of orientational order of these microscopic units. The fluctuating orientational order is systematically described by fluctuating moments of the orientational distribution function, starting with its dipole moment characterizing polar orientational order, quadrupole moment describing the more common nematic orientational order, octupole moment describing tetrahedratic order \cite{fel1995} and so on, in principle.

Governed by symmetry, orientational order is generally coupled to other system variables or external fields. However, in an isotropic system such effects are macroscopically significant only when the system is in the vicinity of an orientational phase transition (if it exists) like the transition from the isotropic to the nematic phase, where orientational fluctuations become large and orientational order gets more susceptible to the influence of external and other internal variables. 

In systems featuring a reduction of microscopic degrees of freedom (e.g., the connectivity of a polymer chain presents a microscopic constraint reducing the configurational space of the monomers in comparison with non-polymerized monomers), however, the coupling of the orientational order to the displacement field, in particular to density variations, is a geometrical necessity and is not related to the proximity of an orientational phase transition threshold or its very existence. Moreover, being geometrical (unbreakable) rather than energetic, such constraint is inherently robust and is hardly affected by any system variables except those that enter the constraint explicitly. As such, the response dictated by this constraint is well-defined, universal and remains unaltered when the system traverses its path in the parameter space.

It has been recognized several decades ago that a microscopic geometrical constraint in the form of a line in the so-called {\it line liquids} \cite{nelson1991,kamien1993,nelson2002}, which include magnetic flux lines as a vortex liquid in type II superconductors, chains of particles in ferro- and electrorheological fluids formed in an external field, and chains of connected monomers in main-chain polymer nematics, comes out on the macroscopic level as a constraint on continuum fields used to describe the coarse-grained configuration of such lines. That is, even if we do not have access to probing the microscopic structure directly, we can nevertheless detect the presence of the microscopic constraint through a macroscopic response of the system.

\begin{figure}[h]
  \centering
    \includegraphics[width=86mm]{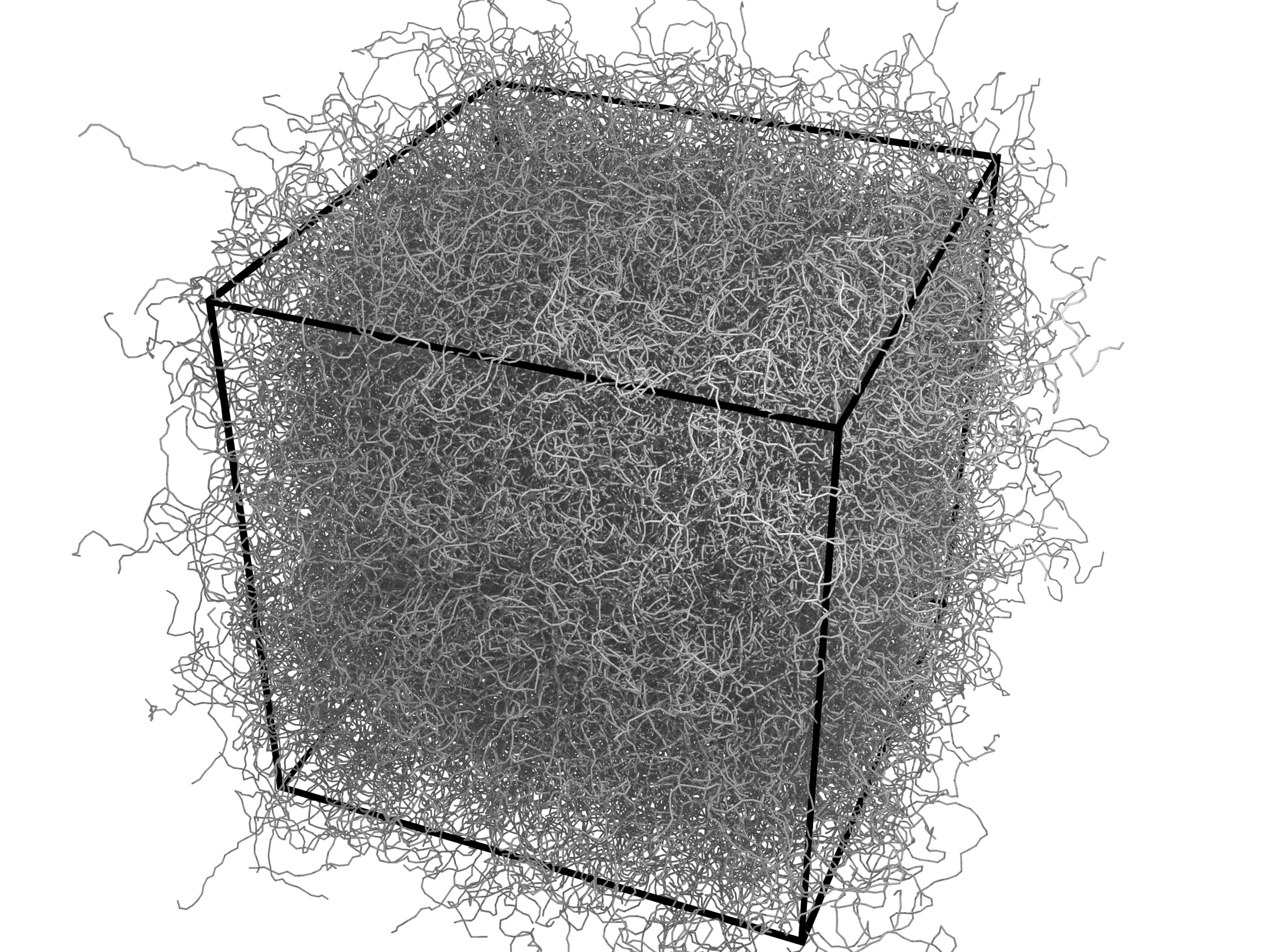}\\[1mm]
	\includegraphics[width=70mm]{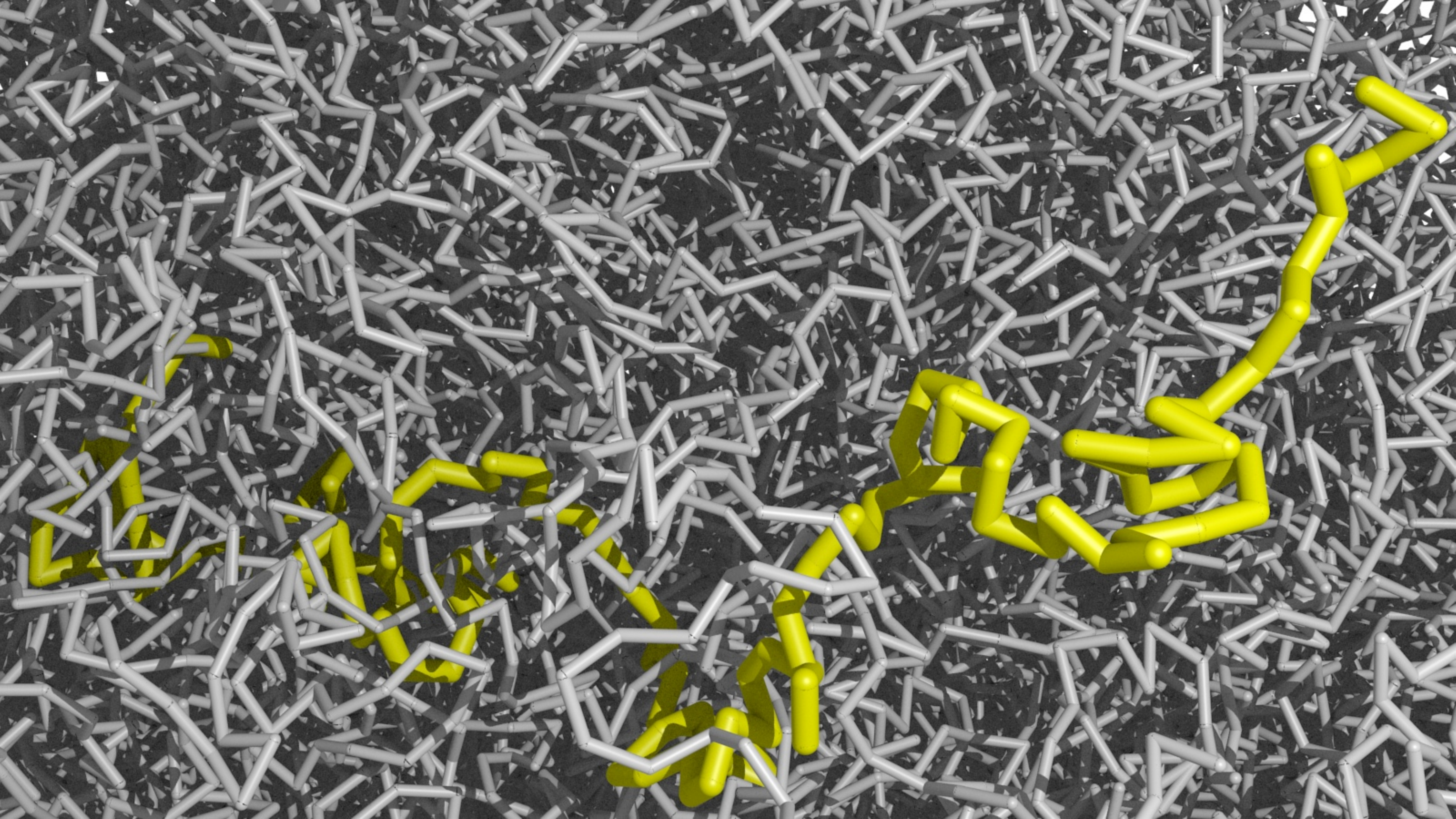}\\[1mm]
    \includegraphics[width=70mm]{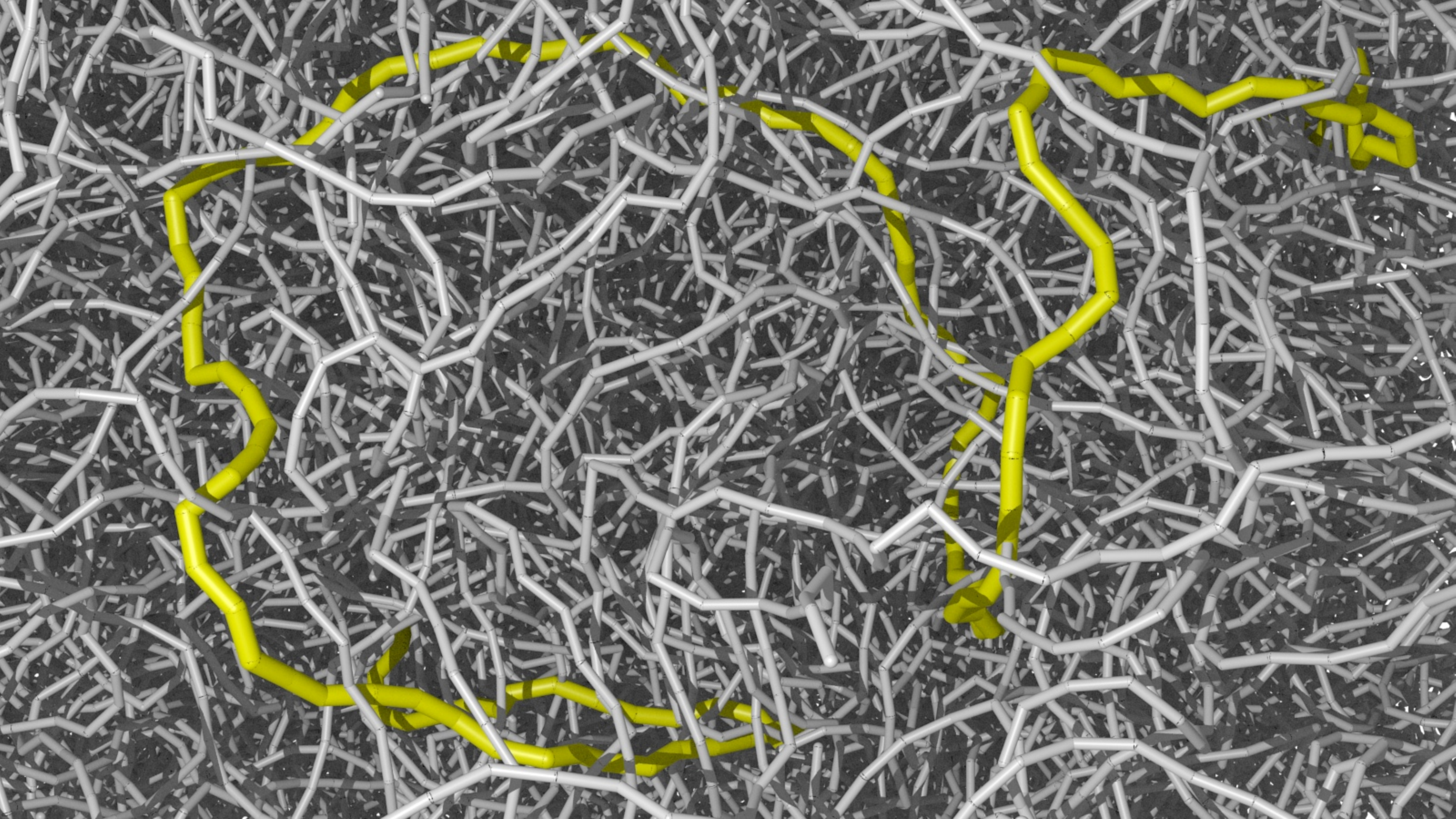}
    \caption{
	Top: snapshot of the simulated isotropic polymer melt with $2^{18}$ monomers (2048 chains with  $N_{\mathrm{s}}=128$ segments) employing periodic boundary conditions.
    Assuming for the monomer diameter the distance at which the repulsive potential is equal to $\kT$,
    we estimate the polymer volume fraction to $\approx 0.11$, corresponding to a polymer melt.
    The close-ups highlight a single chain in the ideally flexible ($\epsilon=0$, middle) and semiflexible ($\epsilon=4.926\,\kT$, bottom) case. 
    }
\label{fig:splay}
\end{figure}

In linear (main-chain) nematic polymers, the continuity constraint was derived in the form of the so-called vectorial conservation law for the ``polymer current'' density \cite{svensek2013} that incorporates the full orientational order vector ${\bf a}({\bf r})$, $|{\bf a}|\le 1$, as: $~\nabla\cdot(\rho l_0{\bf a}) = \rho^\pm$, where $\rho({\bf r})$ is the volume density of arbitrary segments (e.g.\ monomers) of length $l_0$. Moreover, it is clear by construction \cite{popadic2018} that in this conservation law, ${\bf a}({\bf r})=\langle {\bf t}\rangle$ is exactly the coarse-grained (mesoscopic) polar orientational order of polymer chain tangents $\bf t$. This generalizes the previously proposed \cite{degennes1976, meyer1982, vroege1988,ao1991,selinger1991, doussal1991, nelson1991, kamien1992,selinger1992} Meyer-de Gennes continuity constraint $\nabla\cdot(\rho_s {\bf n}) = \rho^{\pm}$, coupling splay deformation of the nematic director ${\bf n}({\bf r})$ with variations of areal density of chains $\rho_s({\bf r})$ (i.e., ``geometrical'' flux density of the chains, $\rho_s=\rho l_0 |{\bf a}|$).  
The consequence of the continuity constraint is that the density of long chains decreases as they are splayed, since there are not many chain ends available that could populate the so-created voids between the chains. Shorter chains provide more ends, which can fill the voids more easily.  This distinction is captured by the volume density $\rho^\pm({\bf r})$ of chain beginnings ($^+$) and endings ($^-$), which then acts as a source in the continuity equation.

In a homogeneous isotropic system, we have $\rho = \rho_0$, ${\bf a} = 0$, and $\rho^\pm=0$ in equilibrium. Any deviation $\delta\rho({\bf r})$, $\delta{\bf a}({\bf r})$ must satisfy the vectorial conservation law constraint 
to the lowest order as: 
	$\rho_0 l_0\nabla\cdot\delta{\bf a} = \delta\rho^\pm$,
that does not involve the density variation, which is thus unaffected by the vectorial constraint. In the isotropic phase, fluctuations of the density and fluctuations of the polar orientational order are thus independent of each other.

Nematic (quadrupolar) orientational order is, however, described by the traceless nematic order tensor $Q_{ij}({\bf r}) = {3\over 2}(\langle t_i t_j\rangle-{1\over 3}\delta_{ij})$, where for linear polymers the averaging is again over chain tangents in a mesoscopic volume centered at $\bf r$.
As shown \cite{svensek2013,svensek2016}, a rigorous conservation law can be derived not only for the polar order, 
but also for the quadrupolar order of the chains, which is, unlike polar order, insensitive to chain backfolding. 
A completely general form of this tensorial conservation law (see Appendix \ref{sec:tensorial} for formal derivation) for arbitrary number of chains with arbitrary length and flexibility is
\begin{equation}
	\partial_j\left[\rho(Q_{ij}+{\textstyle{1\over 2}}\delta_{ij})\right] = {\textstyle{3\over 2}{1\over l_0}}g_i + {\textstyle{3\over 2}}\rho k_i,  
	\label{tensorial}
\end{equation}
where the volume density ${\bf g}({\bf r})$ of chain end tangents, defined as pointing inwards, and the average chain curvature vector ${\bf k}({\bf r})$, multiplied by the density, both play the role of the sources in this continuity equation and can be furthermore considered as independent for sufficiently long chains. 
%
The average chain curvature vector source reflects the effect of the chain folds, which can fill the voids created by splay in a similar way as chain ends do. 
The stiffer and longer the chains, the more expensive are the sources and the stronger is the constraint. Similarly to the vectorial case, 
the tensorial analogue
Eq.~(\ref{tensorial}) is an exact macroscopic implication of the microscopic polymer chain connectivity.

We have theoretically and numerically shown \cite{popadic2018} that it is possible to amend the vectorial continuity equation 
by introducing the ``recovered'' polar order \cite{svensek2016,popadic2018}, such that it can be rigorously applied to the {\it uniaxial} nematic phase with general chain backfolding.\footnote{In the orientationally ordered, uniaxial nematic phase (not considered here), the sources of the tensorial conservation law Eq.~(\ref{tensorial}), projected parallel to the nematic direction $\bf n$, are analogous to those of the vectorial conservation law for the ``recovered'' polar order \cite{svensek2016,popadic2018}: the density of chain end tangents ${\bf g}\cdot{\bf n}$ corresponds to the scalar density of chain ends, while the average curvature vector density $\rho l_0{\bf k}\cdot{\bf n}$ corresponds to the scalar density of backfoldings. The projection of the tensorial constraint parallel to $\bf n$ is analogous (but not identical) to the vectorial constraint for the recovered polar order. The two constraints become identical \cite{svensek2016} only in the limit of perfect orientational order.}
Nevertheless, a complete description of the nematic phase should be based on the full nematic $\sf Q$-tensor and the corresponding continuity constraint Eq.~(\ref{tensorial}), which is needed if biaxiality is important or topological (half-integer) defects of the nematic phase are considered.
In the isotropic phase, however, nematic fluctuations are inherently tensorial and {\it cannot} be described otherwise than with the $\sf Q$-tensor.
Therefore, the tensorial continuity equation must be inevitably used in this case --- 
a situation, which has not been hitherto considered.

In equilibrium, $Q_{ij}=0$ and a deviation $\delta Q_{ij}$ must satisfy the constraint Eq.~(\ref{tensorial}), that to the lowest order yields
\begin{equation}
	\rho_0\, \partial_j \delta Q_{ij} + {\textstyle{1\over 2}}\partial_i\delta\rho = {\textstyle{3\over 2}{1\over l_0}}\delta g_i + {\textstyle{3\over 2}} \rho_0 \delta k_i.
	\label{tensorial_fluct}
\end{equation}
Unlike the fluctuations of polar ordering,
the fluctuations of nematic (quadrupolar) ordering and density are generally coupled in first order even in an orientationally disordered, isotropic phase, Fig.~\ref{fig:splay}. This situation is thus quite generic and applies in principle to any linear polymer melt/solution like polyethylene, polyvinyls, polyamides, polyesters, polystyrene, polycarbonates etc.
The nematic order--density coupling is particularly strong in the case of long and stiff polymer chains, since there the fluctuations of the r.h.s.\ of Eq.~(\ref{tensorial_fluct}) are costlier and thus weaker. But as we will show, it can be significant already for chains as short as a couple of units.

Thus, the tensorial constraint has profound implications already for the isotropic phase.
A quick inspection of Eq.~(\ref{tensorial_fluct}) shows that a fluctuation with the wave vector in $z$ direction (keeping in mind that in the isotropic system all directions are equivalent) couples $\delta\rho$ and $\delta Q_{zz}$. Hence, a spatial variation of density or concentration will induce nematic order and thereby optical anisotropy in an otherwise isotropic system. We will show that this constraint gives rise to a class of interesting macroscopic phenomena in linear polymers like acoustic birefringence in polymer melts or osmotic birefringence in polymer solutions.

Moreover, we also present a methodology that enables one to determine the parameters of the continuum description of the isotropic polymeric liquid employing microscopic simulations. This bears some resemblance to well-established derivations of coarse-grained potentials in molecular simulations, e.g., coarse-graining with the relative entropy \cite{shell2008,noid2013,foley2015}. By using known atomistic or coarse-grained force fields of specific linear polymers, it is possible to accurately extract from numerically calculated correlation functions realistic values of the macroscopic parameters, that correspond to the specific polymeric material. In this light, microscopic simulations of dense phases of DNA \cite{zavadlav2017,podgornik2018} can fix the parameters of coarse-grained, continuum descriptions used e.g.\ to study the packing and ordering properties of nano-confined DNA as in e.g.\ viral capsids \cite{svensek2010,svensek2012} or nanochannels \cite{reisner2005,gupta2015}.

\section{Correlation functions of collective fluctuations}
\label{sec: correlations}

\noindent
We first present the results for fluctuations of the isotropic linear polymer system in the continuum description.
A minimal free-energy density of the isotropic phase taking into account density variations, nematic fluctuations satisfying $\delta Q_{kk}=0$ by definition, and the constraint Eq.~(\ref{tensorial_fluct}),  
is
\begin{eqnarray}
	f &=& {1\over 2}B\left({\delta\rho\over\rho_0}\right)^2 
    + {1\over 2}B'\left({\partial_i\delta\rho\over\rho_0}\right)^2 \label{f} \\ 
    &+& {1\over 2} A \left(\delta Q_{ij}\right)^2 + {1\over 2} L \left(\partial_k\delta Q_{ij}\right)^2\nonumber\\
&+& {1\over 2}G\left({\textstyle{2\over 3}} \rho_0 l_0\right)^2\left[\partial_j\delta Q_{ij}+{\textstyle{1\over 2}}\partial_i\left({\delta\rho\over\rho_0}\right)\right]^2, \nonumber    
\end{eqnarray}
where $B$ is the bulk modulus, $\rho_0$ is the volume number density of monomers, $A$ is the ``nematic order stiffness'' and $B'$ and $L$ (the nematic elastic constant) are penalizing $\rho$ and $\sf Q$ gradients. The density and nematic correlation lengths are $\xi_\rho\sim\sqrt{B'/B}$ and $\xi\sim\sqrt{L/A}$, respectively. 
The constraint due to the tensorial conservation law Eq.~(\ref{tensorial_fluct}) is taken into account by a quadratic potential penalizing its sources, where $G\left({\textstyle{2\over 3}} \rho_0 l_0\right)^2\equiv\tilde{G}$ is the strength of the constraint. A minimal model for $G$ with the final result Eq.~(\ref{G_combined}) is developed in Sec.~\ref{sec:sources}.

In Fourier space, 
$u({\bf q}) = \int\!{\rm d}^3r\, u({\bf r}){\rm e}^{-{\rm i}{\bf q}\cdot{\bf r}}$, and with $\delta\rho/\rho_0\equiv \delta\tilde\rho$,
the free-energy density Eq.~(\ref{f}) is
\begin{eqnarray}
	f({\bf q}) &=& {1\over 2}(B+B'q^2) |\delta\tilde\rho|^2 + {1\over 2} (A+L q^2) |\delta Q_{ij}|^2\nonumber\\
    &+& {1\over 2}\tilde{G}\left|q_j\delta Q_{ij}+{\textstyle{1\over 2}}q_i\delta\tilde\rho\right|^2.
    \label{fq}
\end{eqnarray}
The free energy is $F = \int\!{\rm d}^3 r f = (1/V)\sum_{\bf q} f({\bf q})$, where $V$ is the volume of the system. By equipartition, the energy corresponding to an individual quadratic contribution $f_{i}({\bf q})$ to Eq.~(\ref{fq}) is
	$\langle f_{i}({\bf q})\rangle/V = \kT/2$, with	
$k_{\rm B}$ the Boltzmann constant and $T$ the temperature.

To determine the fluctuation amplitudes,
the quadratic form Eq.~(\ref{fq}) is diagonalized.
Since the system is isotropic, without loss of generality we may assume ${\bf q} = q \hat{\bf e}_z$, where $z$ is an arbitrarily chosen direction defining the $z$ axis of the coordinate system. Axes $x$ and $y$ are then fixed arbitrarily and all results at a given $q$ must be invariant to rotations of the tensors in the $xy$ plane. 
Since ${\mathsf Q}$ is traceless by definition, only two of $\delta Q_{xx}$, $\delta Q_{yy}$, and $\delta Q_{zz}$ are independent. Conforming to the symmetry of the problem, we put $\delta Q_{zz} = -(\delta Q_{xx}+\delta Q_{yy})$ and take $\delta Q_{xx}$, $\delta Q_{yy}$ as the variables. Moreover, 
for the remaining three variables we take $\delta Q_{xy}$, $\delta Q_{xz}$, $\delta Q_{yz}$, which represent also their transposes and will thus give twofold free-energy contributions. 
With that, 
the diagonalized free-energy form Eq.~(\ref{fq}) is
\begin{eqnarray}
	f(q) &=& {1\over 2}(A+L q^2) \left(2|\delta Q_{xy}|^2 + \left|\delta Q_{xx}-\delta Q_{yy}\right|^2\right)\label{fqdiag}\\
	&+& \left[A+(L+{\textstyle{1\over 2}}\tilde{G})q^2\right]\left(|\delta Q_{xz}|^2+|\delta Q_{yz}|^2\right)\nonumber\\
	&+& {\lambda^+\over v_+^2}\left|a_+\delta\tilde\rho+\delta Q_{zz}\right|^2 + 
		 {\lambda^-\over v_-^2}\left|a_-\delta\tilde\rho+\delta Q_{zz}\right|^2,\nonumber
\end{eqnarray}
where we reverted to $\delta Q_{zz}$ in the last two terms. The expressions $v^2_\pm=2+a^2_\pm$, $\lambda^\pm$ and $a_\pm$ are real and are given in Appendix \ref{sec:diagonalization}.

The autocorrelations of the variables that appear quadratically in Eq.~(\ref{fqdiag}) follow immediately from equipartition, Eqs.~(\ref{Qxyfluct}) and (\ref{Qxzfluct}).
The fluctuation $\delta Q_{xx}-\delta Q_{yy}$ leaves $\delta Q_{zz}$ unaltered and its free-energy cost is the same as that of $\delta Q_{xy}$ (note the twofold contribution of this latter off-diagonal term) --- as it must be to recover the isotropy in the $xy$ plane. As such, it does not bring anything new.
The last two terms in Eq.~(\ref{fqdiag}) represent the contributions of two coupled fluctuation modes, i.e., it is just the component $\delta Q_{zz}$ (and thus also the sum $\delta Q_{xx}+\delta Q_{yy}$) that is coupled to density. 

The complete collection of Fourier-component thermodynamic correlations in space (see Appendix \ref{sec:diagonalization} for details) is
\begin{eqnarray}
	&&{1\over N_0}\langle|\delta Q_{xy}|^2\rangle = {\kT\over 2}{1\over\rho_0}{1\over A + Lq^2}\label{Qxyfluct}\\
	&&{1\over N_0}\langle|\delta Q_{\{xz,yz\}}|^2\rangle = {\kT\over 2}{1\over\rho_0}{1\over A+(L+{\textstyle{1\over 2}}\tilde{G})q^2}\label{Qxzfluct}\\
	&&{1\over N_0}\langle|\delta Q_{zz}|^2\rangle = {\kT\over 2}{4\over\rho_0}\left[3A+\left(3L+{8\tilde G \tilde B\over 4\tilde B+\tilde G q^2}\right)q^2\right]^{-1}\label{Qzzfluct}\\
	&&{1\over N_0}\langle|\delta\tilde\rho|^2\rangle = {\kT\over 2}{1\over\rho_0}8\left[4\tilde B+{3\tilde G(A+Lq^2)q^2\over 3A+(3L+2\tilde G)q^2}\right]^{-1}\label{rhofluct}
\end{eqnarray}
\begin{eqnarray}
    &&{1\over 2N_0}\langle\delta\tilde\rho^*\, \delta Q_{zz}+\delta\tilde\rho\, \delta Q_{zz}^*\rangle = -{\kT\over 2}{1\over\rho_0}\times \label{cross-correlation} \\&&~~\times {8\tilde G q^2\over 12 A\tilde B+[12\tilde B L+(3A+8\tilde B)\tilde G]q^2+3\tilde G L q^4},\nonumber
\end{eqnarray}    
where $\tilde B = B+B'q^2$. 
It could be seen already by inspection of Eq.~(\ref{fq}) that $\delta Q_{xy}$, $\delta Q_{xz}$, and $\delta Q_{yz}$ are diagonal and the free-energy cost of $\delta Q_{xy}$ is different from that of $\delta Q_{xz}$ and $\delta Q_{yz}$, which is indeed confirmed by the result Eqs.~(\ref{Qxyfluct})-(\ref{Qxzfluct}). Importantly, since the system is isotropic there is no elastic anisotropy and thus this difference is a signature of the tensorial constraint alone.
From the fluctuations Eqs.~(\ref{Qxyfluct})-(\ref{Qxzfluct}) one can efficiently determine the values of the parameters $A$, $L$ and the coupling strength $\tilde G$, which we will make use of in the following. 

The coupling of the fluctuations $\delta Q_{zz}$ and $\delta\tilde\rho$, seen already from Eq.~(\ref{fqdiag}), is reflected in a nonzero cross-correlation Eq.~(\ref{cross-correlation}).
This negative correlation is again the signature of the tensorial constraint and vanishes in the absence of the constraint when $\delta\tilde\rho$ and $\delta Q_{zz}$ are decoupled.

\section{Comparison with microscopic simulations}

\noindent
To affirm the existence of the tensorial constraint, we employ Monte Carlo (MC) simulations 
of discrete worm-like chains (WLC), and compare the simulated correlation functions of static long wavelength fluctuations with Eqs.~(\ref{Qxyfluct})-(\ref{cross-correlation}) following from the continuum theory. 
Validating the predictions of the macroscopic theory with molecular-level computer simulations of polymers \cite{dellesite2017,praprotnik2008,binder2018} is challenging, since such simulations must i) address the long-wavelength limit and ii) realize different regimes of chain backfolding (hairpin formation). Thus, it is essential to consider large systems of preferably long polymer chains \cite{kremer1990}, where the sampling must include statistically independent (decorrelated) configurations.
We fulfill these requirements benefiting from a recently developed mesoscopic model \cite{gemunden2015,popadic2018} of discrete WLCs.
The modeled system contains $N_{\rm c}$ WLCs comprised of $N_{\rm s}$ linearly connected segments of fixed length $l_0$. Consecutive
segments are subjected to a standard angular potential Eq.~(\ref{angular_potential}) with strength $\epsilon$ controlling the WLC bending stiffness. Moreover, all segments possess a non-bonded isotropic repulsive interaction with finite microscopic range $2 l_0$ and strength $\kappa$. 
See Appendix \ref{sec:simulation} and Ref.~\cite{popadic2018} for details of the numerical method and the simulated mesoscopic WLC model.

We study large isotropic melt systems containing 
$N_0 = N_{\rm c}N_{\rm s} = 2^{18}$ segments.
An example of the simulation snapshot is shown in Fig.~\ref{fig:splay} (top).
The configurations are equilibrated through MC starting from a nematic phase, with the chains stretched along the $z$ axis of the laboratory frame and their centers of mass randomly distributed in a cubic box with periodic boundary conditions.
The MC algorithm utilizes the standard~\cite{frenkel2001,tuckerman2010} slithering-snake moves, as well as volume fluctuation moves at
pressure $Pl_0^3/(k_{B}T) = 2.87$ resulting in simulation box sides of length $\langle L \rangle/l_0 \sim 66$ and system's volume fluctuations of $\sim 0.1\%$.
While working in the isothermal-isobaric ensemble is computationally more expensive, it is useful when determining the bulk modulus.
The efficient soft model enables us to accumulate large sequences of statistically decorrelated isotropic configurations, which allow for direct validation 
of the macroscopic theory via the correlations Eqs.~(\ref{Qxyfluct})-(\ref{cross-correlation}) (see Appendix \ref{sec:simulation} for details of their extraction from simulation data).

\begin{figure}[h]
  \centering
    \includegraphics[]{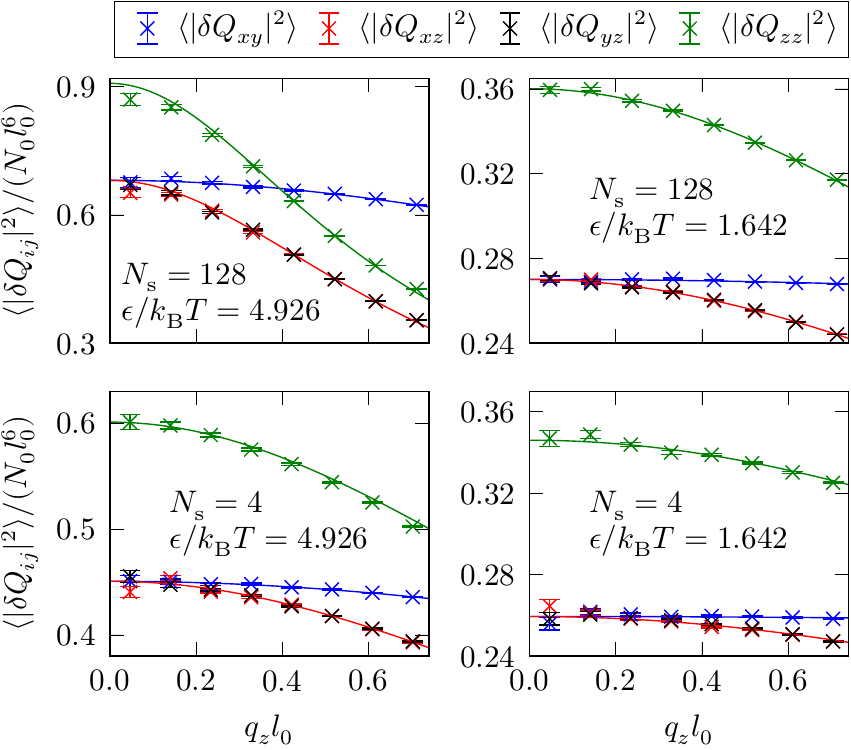}
    \caption{Nematic ($\delta Q_{ij}$) fluctuations of stiffer ($\epsilon=4.926\, \kT$, left) and more flexible ($\epsilon=1.642\,\kT$, right) chains with $N_{\mathrm{s}}=128$ (top) and $N_{\mathrm{s}}=4$ (bottom) segments, fitted with Eqs.~(\ref{Qxyfluct}) and (\ref{Qxzfluct}) to extract the values of the parameters $A$, $L$ and $\tilde G$.
    The $\langle|\delta Q_{zz}|^2\rangle$ curves are not fitted but are direct plots of the theoretical result Eq.~(\ref{Qzzfluct}).
    }
\label{fig:QijQij}
\end{figure}

Fig.~\ref{fig:QijQij} shows examples of normalized fluctuations $\langle|\delta Q_{ij}|^2\rangle$ of the nematic ordering, calculated in simulations, for different chain lengths (top, bottom) and different chain flexibilities (left, right).   
While $\langle|\delta Q_{xz}|^2\rangle$ and $\langle|\delta Q_{yz}|^2\rangle$ coincide, it is confirmed that they are different from the fluctuations $\langle|\delta Q_{xy}|^2\rangle$, in accord with the results Eqs.~(\ref{Qxyfluct})-(\ref{Qxzfluct}). The difference grows with the strength $\tilde G$ of the tensorial constraint, increasing with length ($N_{\mathrm{s}}$) and bending stiffness ($\epsilon$) of the chains. The $\langle|\delta Q_{xy}|^2\rangle$ points are fitted with Eq.~(\ref{Qxyfluct}), determining the parameters $A$ and $L$. With these parameters fixed, the $\langle|\delta Q_{xz}|^2\rangle$ and $\langle|\delta Q_{yz}|^2\rangle$ data are then fitted with Eq.~(\ref{Qxzfluct}) and the strength $\tilde G$ is determined. 

It would be natural to determine the modulus $B$ from the density autocorrelation (the structure factor) Eq.~(\ref{rhofluct}). It turns out, however, that the theoretical Lorentzian profile of the structure factor in the continuum picture, Eq.~(\ref{rhofluct}), is completely overridden by the influence of the discrete structure of the simulated WLCs even for the lowest $q$'s that we can achieve with our simulation box size. 
Therefore, we put $B'$ to zero and determine $B$ from fluctuations $\delta V$ of the simulation box volume $V_0$,  $\langle\delta V^2\rangle = V_0 \kT/B$.
This agrees with the structure factor Eq.~(\ref{rhofluct}) for $q=0$, recalling that for a homogeneous density variation, $\delta V/V_0 = -\delta\rho/\rho_0$ and the Fourier component is $\delta\rho(q=0) = V_0\delta\rho$.

\begin{figure}[h]
  \centering
    \hspace{-6mm}\includegraphics[width=9cm]{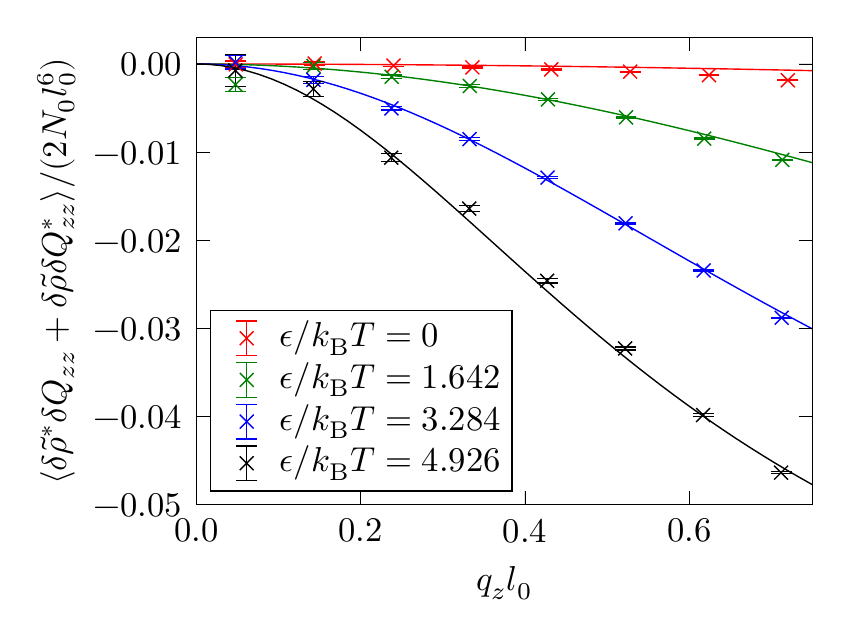}\\
    \includegraphics[]{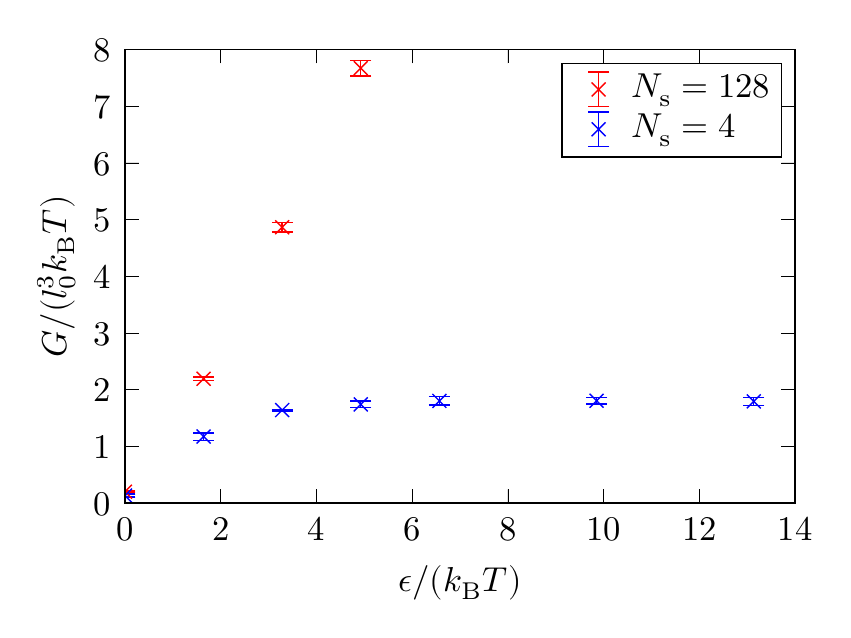}
    \caption{Top: dimensionless density--nematic cross-correlations for chains with length $N_{\mathrm{s}}=128$ and varying flexibility. The curves are plots of Eq.~(\ref{cross-correlation}) (no fitting). Bottom: dimensionless coupling strength $G$ vs.\ bending stiffness $\epsilon/(\kT)$, determined from fits of the numerically calculated fluctuations $\langle |\delta Q_{xy}|^2\rangle$, $\langle |\delta Q_{xz}|^2\rangle$, $\langle |\delta Q_{yz}|^2\rangle$, Fig.~\ref{fig:QijQij}.
}
\label{fig:Qzzdens(eps)}
\end{figure}

A direct display of the density--nematic coupling is the cross-correlation Eq.~(\ref{cross-correlation}) between $\delta\tilde\rho$ and $\delta Q_{zz}$, in theory directly proportional to the strength of the coupling $\tilde G$. In Fig.~\ref{fig:Qzzdens(eps)} (top), the cross-correlation is shown for several chain flexibilities. Here, the theoretical curves are not fitted, but correspond to the prediction Eq.~(\ref{cross-correlation}), using the values of the parameters $A$, $L$, $\tilde G$ and $B$ extracted from the fluctuations $\delta Q_{xy}$, $\delta Q_{xz}$, $\delta Q_{yz}$ (Fig.~\ref{fig:QijQij}) and $\delta V$. The same is valid for the $\langle|\delta Q_{zz}|^2\rangle$ curves in Fig.~\ref{fig:QijQij}, which are plots of Eq.~(\ref{Qzzfluct}) with the same parameter values. For sufficiently long chains, the coupling strength $G$ extracted from the simulation data, Fig.~\ref{fig:Qzzdens(eps)} (bottom), clearly increases with chain stiffness, as anticipated from the fact that the fluctuations $\delta{\bf k}$ of the curvature source in Eq.~(\ref{tensorial_fluct}) get costlier. In the case of very short chains, $G$ becomes saturated already at lower stiffness, since in this case the source $\delta{\bf g}$, corresponding to the density of chain end tangents, is dominant (next Section reveals that this is only a partial reason).

%
%

Fig.~\ref{fig:Qzzdens(eps)} thus presents direct evidence of the connection between density variations and the emergence of nematic orientational order in otherwise isotropic polymeric liquid, which is also well described by the theoretical cross-correlation Eq.~(\ref{cross-correlation}). Moreover, the strength of this coupling increases for long and stiff chains, as expected and empirically confirmed in Fig.~\ref{fig:Qzzdens(eps)} (bottom).

\section{Theoretical model of the sources}
\label{sec:sources}

\noindent
In this additional step, we build a theoretical model to predict the coupling strength $G$ on the basis of length and flexibility of the chains. Similar to what has been done in Ref.~\cite{popadic2018}, we resort to a minimal model of the sources of the continuity equation Eq.~(\ref{tensorial}), in the sense that i) we treat both macroscopic sources as composed of independent microscopic contributions to chain-end tangent density $\bf g$ (Sec.~\ref{sec:ends}) and average chain curvature ${\bf k}$ (Sec.~\ref{sec:curvature}), respectively, and ii) we combine both sources into a single unified source ${\bf h} = {\bf g}+\rho_0 l_0{\bf k}$ with a properly weighted relative composition (Sec.~\ref{sec:combined}). Only this latter case allows the constraint Eq.~(\ref{tensorial_fluct}) to be taken into account simply by a penalty potential term in the free-energy density Eq.~(\ref{f}), which means that no additional variables for the sources are required.

Thus, we shall construct the nonequilibrium free energy cost of the sources of Eq.~(\ref{tensorial}) in the simplest possible way and in lowest, quadratic order of the sources. 
Since ${\bf g}=0$ in equilibrium, the variation $\delta\bf g$ of the density of end tangents involves only the variation of their average orientation, and does not involve the variation of their density.
(This is true also in the nematic phase.)
Similarly, the equilibrium average chain curvature $\bf k$ is zero, so the variation of $\rho{\bf k}$ does not involve the variation of the density, as already made explicit in Eq.~(\ref{tensorial_fluct}). 
(This is true also in the nematic phase, provided it is not bent.)

\subsection{End tangents}
\label{sec:ends}

\noindent
We construct a purely entropic nonequilibrium orientational free energy of orientationally independent end tangents, taking into account their dipolar ordering that results in nonzero $\bf g$. 
The orientational part of the entropic free energy is then
\begin{eqnarray}
	F(p_1) = -TS(p_1) &=& \kT \int\! {\rm d}\Omega\, p(\Omega) \ln{p(\Omega)},
    \label{bgdjks}
\end{eqnarray}
where  $p(\Omega) = p_0 + p_1 \sqrt{\textstyle{3/ 2}}P_1(\cos\theta)$ is the orientational distribution function of end tangents with respect to the solid angle $\Omega$, 
$P_1(\cos\theta)=\cos\theta$ is the first Legendre polynomial and $p_0 = 1/2$ is fixed by the normalization  $\int_{-1}^1{\rm d}(\cos\theta)p(\Omega) = 1$. In the isotropic system, the orientation of the $z$ axis of the spherical coordinate system is arbitrary. Moreover, 
$p_1 = \sqrt{\textstyle{3/ 2}}\langle\cos\theta\rangle$
is the dipole moment of end tangent orientations (a nonzero $p_1$ means a nonzero $\bf g$), which is the only parameter of the orientational distribution. 

One can verify that the first derivative ${{\rm d}F\over{\rm d}p_1}\big\vert_{p_1=0}=0$, so that the free energy is indeed minimum for the isotropic orientational distribution of end tangents.
Hence, for a single chain end, we have
\begin{equation}
	\Delta F(p_1) = {\frac12}{{\rm d}^2F\over{\rm d}p_1^2}\big\vert_{p_1=0}~p_1^2 = {\frac12} 2\kT\, p_1^2.
\end{equation}

For many independent chain ends with density $\rho_0^\pm$, the free-energy density is thus
\begin{equation}
	\Delta f = {1\over 2}\,2\rho_0^\pm \kT\, p_1^2 = {1\over 2} {3\kT\over \rho_0^\pm}\, g^2.
    \label{g_cost}
\end{equation}
where we took into account that with respect to a given direction $g = \rho_0^\pm\langle\cos\theta\rangle = \rho_0^\pm\sqrt{\textstyle{2/ 3}}p_1$.
This is thus the entropic free energy cost of the density $g=|{\bf g}|$ of independent tangents of chain ends with number density $\rho_0^\pm$. Not unexpectedly, it has the same $1/\rho_0^\pm$ dependence as the entropic free-energy density of the source of the vectorial constraint \cite{popadic2018}.

\subsection{Chain curvature}
\label{sec:curvature}

\noindent
We shall construct a nonequilibrium free-energy density of nonzero local average curvature of the chains.
The bending free energy of a single WLC with nearest-neighbour bending interactions (which also correponds to the angular potential Eq.~(\ref{angular_potential}) used in the microscopic simulation) is $\Delta F = \sum_i \Delta F_i$, with
\begin{equation}
 	\Delta F_i = {1\over 2}\epsilon l_0^2|{\bf k}^i|^2 = \epsilon \left( 1 -\cos{\theta_{i,i+1}}\right),
    \label{F(theta)}
\end{equation}
where ${\bf k}^i = ({\bf u}^{i+1}-{\bf u}^i)/l_0$, ${\bf u}^i$ is the unit vector along the $i$-th segment of the chain, and $\cos\theta_{i,i+1} = {\bf u}^i\cdot{\bf u}^{i+1}$.
Thus, considering only the bending free energy Eq.~(\ref{F(theta)}), the individual microscopic curvature elements, corresponding to individual monomer joints, are independent. The relevance of this independent joint assumption in the actual system including also non-bonded interactions is demonstrated in Appendix \ref{sec:bend}.
If the chain segments are bent only slightly, i.e., when $\kT\ll \epsilon$, then the two components of ${\bf k}^i$ can be as usual considered 
ranging from $-\infty$ to $\infty$, such that equipartition holds and
\begin{equation}
	\langle (k^i_{1})^2\rangle = {\kT\over \epsilon l_0^2},
    \label{k01}
\end{equation}
where $k^i_{1}$ is one of the components.
In the continuum limit, ${\bf k}(s) = {\rm d}{\bf t}/{\rm d}s$ is the local chain curvature vector and the free energy is $\Delta F = {\textstyle\frac12}K \int\! {\rm d}s\, k(s)^2$, where $s$ is the arclength along the chain and $K=\epsilon l_0$ is the bending rigidity of the continuous WLC, if one disregards the rather small influence of the non-bonded interactions on the flexibility of the discrete WLC.

To arrive at the free energy of the collective (average) chain curvature $\bf k$, one has to find the configuration of ${\bf k}^i$'s that corresponds to the most complete equilibrium \cite[pp.\,335, 398]{landau1980} at a given $\bf k$, e.g., for segment pairs that are on average perpendicular to $\bf k$ this would simply mean ${\bf k}^i = {\bf k}$.
In the isotropic system, however, the segments point in all directions. Let ${\bf k}=k\hat{\bf e}_z$. By symmetry, the segment pairs oriented along $z$, i.e., $\theta=0$, are not bend on average and therefore do not contribute either to $\bf k$ or to the free energy. The contributions of the segment pairs lying in the $xy$ plane ($\theta=\pi/2$) are on the other hand maximum (and equal). We shall assume that the magnitude of the joint's curvature vector ${\bf k}_0 = -k_0\hat{\bf e}_\theta$ goes as $k_0\propto \sin\theta$, such that its contribution to the macroscopic curvature is ${\bf k}_0\cdot\hat{\bf e}_z\propto \sin^2\theta$. Requiring that its solid angle average is $k$, we get
	$k_{0}(\theta) = {3\over 2}k\sin\theta$.
For a constant density of monomers $\rho_0$, the corresponding effective curvature free-energy density is then obtained by averaging Eq.~(\ref{F(theta)}) over the solid angle, 
with the result
\begin{equation}
	\Delta f(k) = {1\over 2}{{3\over 2}}\epsilon l_0^2\rho_0\, k^2 = {1\over 2}{{3\epsilon l_0^2\over 2\rho_0}}\, (\rho_0 k)^2,
    \label{f_K}
\end{equation}
where $k$ is the macroscopic (average) chain curvature in an arbitrary direction of the isotropic system and the quantity $\rho_0{\bf k}$ enters the source of the tensorial continuity equation Eq.~(\ref{tensorial}).


Considering in Eq.~(\ref{f_K}) only one monomer, i.e., $\rho_0=1/V$, we get 
\begin{equation}
	\langle k^2_1\rangle^0 = {2\kT\over 3\epsilon l_0^2} \equiv {2\over 3}{1\over l_0 \xi_p},
    \label{<k2>}
\end{equation}
where the superscript $^0$ stands for $\kT/\epsilon\to 0$.
One can verify that the same result Eq.~(\ref{<k2>}) is obtained directly by averaging the average square of the curvature in a given direction Eq.~(\ref{k01}) over all possible chain orientations, which corroborates the reasoning leading to Eq.~(\ref{f_K}).
In Eq.~(\ref{<k2>}) we have added the general connection between $\langle k^2_1\rangle$ and the persistence length $\xi_p$ of the chain \cite[p.\,399]{landau1980}.

The above developments are approximate and rely on the assumption of small {\it collective} curvatures, which is correct for thermal fluctuations but cannot describe externally imposed arbitrary curvature conditions. In that case nonlinear effects become non-negligible and a more general theory would be needed.

\subsection{Combined sources}
\label{sec:combined}

\noindent
Finally, we establish a model that describes both sources of the continuity constraint Eq.~(\ref{tensorial_fluct}) on a unified basis, with a single variable
\begin{eqnarray}
	{\bf h} = {\bf g} + \rho_0 l_0 {\bf k},
    \label{h}
\end{eqnarray}
analogous to what has been done in Ref.~\cite{popadic2018} for the chain ends and chain backfolds as the sources of the vectorial continuity constraint for the ``recovered'' polar order.
This enables us to predict the strength $G$ of the constraint Eq.~(\ref{tensorial_fluct}), which is enforced simply by the unified source penalty potential in Eq.~(\ref{f}) rather than by introducing additional system variables for the sources.

Following Eqs.~(\ref{g_cost}) and (\ref{f_K}), the total nonequilibrium free-energy density of the sources is
\begin{eqnarray}
	\Delta f({\bf g},{\bf k}) &=& \Delta f({\bf g})+\Delta f({\bf k}) \label{gk_cost}\\ 
    &=&{1\over 2}\left[
    	 {3\kT\over \rho_0^\pm}\, g^2 + 
    	{3\over 2}{\epsilon l_0^2\over\rho_0}\,(\rho_0 k)^2
    \right]. \nonumber
\end{eqnarray}
Considering the combined source $\bf h$, its free-energy density is obtained by averaging Eq.~(\ref{gk_cost}) over all possible realizations Eq.~(\ref{h}) of $\bf h$,
\begin{equation}
 \Delta\bar f({\bf h}) = -(1/V_1)\,{{\rm d}(\ln Z)/{\rm d}\beta},
    \label{f_averaging}
\end{equation}
where $\beta=1/(\kT)$ and the partition function is given as
\begin{equation}
	Z =  \int\!\!\!\int\!{\rm d}^3 g\,{\rm d}^3 k\,
    {\cal P}({\bf g}){\cal P}({\bf k})\, \delta({\bf g} + \rho_0 l_0 {\bf k}-{\bf h}).
    \label{Z}
\end{equation}
Here both ${\cal P}({\bf g})$ and ${\cal P}({\bf k})$ are thermal Boltzmann weights corresponding to energies $V_1\Delta f({\bf g})$ and $V_1 \Delta f({\bf k})$, 
where 
$V_1$ is a coarse-graining volume that does not appear in the final result. To calculate the average Eq.~(\ref{f_averaging}), 
it is thus sufficient to calculate just the integral Eq.~(\ref{Z}), which
is carried out in spherical coordinates with ${\bf h} = h\hat{\bf e}_z$ and ${\bf g}\cdot{\bf h} = g h \cos\theta$. 
The result is
\begin{equation}
	\Delta\bar f({h}) = {3\kT\over 2}\left({1\over V_1} + {h^2\over \rho_0^\pm + 2\kT\rho_0/\epsilon}\right),
\end{equation}
where the first, constant term $3\kT/(2V_1)$ can be omitted --- it arises due to the fact that the state ${\bf h}=0$ can be realized by ${\bf g} = -\rho_0 l_0{\bf k}\ne 0$, which costs energy (i.e., the ground state energy), while the second term is actually the free-energy density Eq.~(\ref{gk_cost}) of the average source, in accord with the property of the Gaussian distribution $\overline{f(h)} = \overline{\Delta f}(0) + \Delta f(\bar h)$.

Thus, the nonequilibrium free-energy density of the total effective source $\bf h$ in arbitrary direction is $\Delta f({\bf h}) = {\textstyle{1\over 2}}G h^2$, where the result of the combined sources model for the coupling strength $G$ that enters Eq.~(\ref{f}) is
\begin{equation}
	G = {3\kT\over \rho_0^\pm + 2 \kT\rho_0/\epsilon}.
    \label{G_combined}
\end{equation}
This expression also explicitly determines the crossover from chain-end- to chain-curvature-dominated strength of the constraint.
Assuming {\it monodisperse} chains with $N_{\rm s}$ monomers, such that $\rho_0^\pm = 2\rho_0/N_{\rm s}$, and using the connection Eq.~(\ref{<k2>}) with the persistence length $\xi_p$, we can rewrite Eq.~(\ref{G_combined}) as 
\begin{equation}
	G = {3\over 2}{\kT\over\rho_0}{1\over 1/N_{\rm s}+l_0/\xi_p}.
    \label{G_xip}
\end{equation}
The crossover takes place at $\xi_p = N_{\rm s} l_0$, i.e., when the persistence length equals the length of the chain.

Hence, for sufficiently long chains, $G$ will be dominated by the curvature source for any finite value of the bending stiffness $\epsilon$, while the contribution of the end tangents will be less significant. In this case $G\approx (3/2)\epsilon/\rho_0 = (3/2)\kT\xi_p/(\rho_0 l_0)$ becomes directly proportional to the bending stiffness,
corresponding to the semiflexible regime of 
severely bent chains. Throughout this regime, which for long chains is very wide, $G$ is dominated by chain curvature.

\begin{figure}[h]
  \centering
    \includegraphics[width=8.5cm]{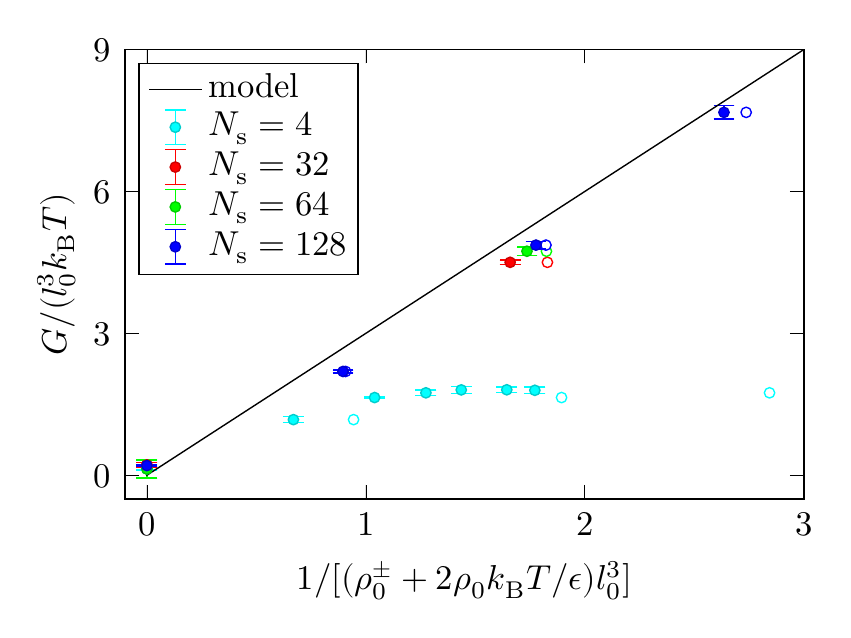}
    \caption{Dimensionless strength of the tensorial constraint $G$, determined in the simulations for several chain lengths $N_{\mathrm{s}}$ and bending stiffnesses $\epsilon$, versus its dimensionless theoretical expression Eq.~(\ref{G_combined}) (solids circles with statistical error bars) --- a direct result without fitted parameters. The contribution of chain ends can be traced with the help of the empty circles, the abscissae of which have $\rho_0^\pm$ put to zero (three rightmost $N_s=4$ open circles fall outside the plot and are not shown).
    }
\label{fig:G(combined sources)}
\end{figure}

Fig.~\ref{fig:G(combined sources)} shows a plot of the dimensionless strength $G$ of the tensorial contraint, determined in the simulations, versus the dimensionless expression Eq.~(\ref{G_combined}).
According to this minimalistic model of combined sources, the points should thus lie on the indicated straight line. For all but shortest chains the agreement is remarkable, all the more so since the theoretical prediction Eq.~(\ref{G_combined}) is direct and involves no fitted parameters.
The problem with very short chains, in particular if they are stiff, is the strong correlation between end tangents of the chain and the in-between profile of its curvature vector, in other words, a rather small number of internal configurational degrees of freedom of the chain. 
Consequently, the sources $\bf g$ and $\bf k$ are not independent, which however was the basic assumption of the combined sources model. 

For the longest chains ($N_{\mathrm{s}}=128$), the two contributions in the denominator of Eq.~(\ref{G_combined}) are, in dimensionless form, 
$\rho_0^\pm l_0^3\sim 0.014$ 
and 
$2\rho_0 l_0^3 \kT/\epsilon \sim  0.37$ (for $\epsilon=4.926\,\kT$, and larger for smaller $\epsilon$).
For the shortest chains ($N_{\mathrm{s}}=4$), the figures are 
$\rho_0^\pm l_0^3\sim 0.44$ 
and 
$2\rho_0 l_0^3 \kT/\epsilon \sim 0.13$ (for $\epsilon=13.136\,\kT$, and larger for smaller $\epsilon$).
That is, in the examples shown in Fig.~\ref{fig:G(combined sources)} the coupling strength $G$ is indeed determined predominantly by chain curvature, except for the stiffer cases of the shortest chains. Regarding the above discussion, this is in accord with Fig.~\ref{fig:splay} (bottom), where one can see that the persistence length, while clearly exceeding the monomer length, is still much shorter than the chain.

Moreover, Fig.~\ref{fig:G(combined sources)} also reveals the relevance of the effective source model: the abscissae of the additional points (empty circles) are obtained by omitting $\rho_0^\pm$ in Eq.~(\ref{G_combined}). Without the contribution of end tangents the agreement is clearly not as good --- the differing slope of the $N_{\rm s}=\{32, 64, 128\}$ triple is particularly noteworthy. The improvement when including $\rho_0^\pm$ is naturally largest for the shortest chains, where the theory however breaks down for the reason mentioned above. 

As exhibited by Fig.~\ref{fig:G(combined sources)}, in all cases the coupling is somewhat weaker than predicted by Eq.~(\ref{G_combined}). 
Plausibly, this slight overall weakening of the constraint is a signature of the fact that the curvature source $\delta{\bf k}$ is not autonomous, but is generally coupled to gradients of $\delta\rho$ and $\delta Q_{ij}$.
A systematic study to quantify all such symmetry-allowed couplings as additions to the free-energy functional Eq.~(\ref{f}) is a natural next step.
In this context, it may seem surprising that the theoretical prediction of the average curvature, Fig.~\ref{fig:1/k12} of Appendix \ref{sec:bend}, which assumes independent pairs of segments, is so accurate. In general, one would expect it to be influenced by these couplings as well. It is, however, physically reasonable that they affect the direction of the curvature fluctuations $\delta{\bf k}$ significantly more than their magnitude. In other words, the magnitude $\langle|\delta {\bf k}|^2\rangle$ is inertly fixed by the free-energy cost Eq.~(\ref{f_K}), whereas the direction of $\delta{\bf k}$ is not distinguished and is therefore prone to other, weaker energy couplings. For increasingly stiff chains, these become inferior compared to the increasing energy cost of the curvature, which in Fig.~\ref{fig:G(combined sources)} seems to be the reason for the improving agreement in the case of stiffer $N_{\rm s}=128$ chains.

\section{Discussion}

\noindent
Once the coupling strength $\tilde G$ has been determined, e.g., by extraction from the fluctuations as shown, also the equilibrium coupling of $\delta\rho$ and $\delta{\sf Q}$ is known. Through this coupling, density or concentration inhomogeneities induce orientational order of the polymer chains, 
which results in a uniaxial dielectric tensor 
$	\varepsilon_{ij} = \varepsilon \delta_{ij}+{\textstyle{2\over 3}}\varepsilon_a  \delta Q_{ij}$
with an anisotropy
$	\varepsilon_{zz}-\varepsilon_\perp = \varepsilon_a \delta Q_{zz}$,
where $\varepsilon$ is the dielectric constant of the isotropic phase and $\varepsilon_a$ the dielectric anisotropy of a phase with perfect orientational order of the chains (maximum possible anisotropy of the material).

The anisotropy of the dielectric constant implies birefringent optical response.
For example, a density (acoustic) plane wave $\delta\tilde\rho({\bf r},t)=\delta\rho({\bf r},t)/\rho_0$ with wave vector ${\bf q}=q\hat{\bf e}_z$ induces uniaxial nematic ordering along $z$, Eq.~(\ref{Qzz_sound}) (see Appendix \ref{sec:birefringence}). The ordering is oblate ($\delta Q_{zz}<0$) in compressions and prolate ($\delta Q_{zz}>0$) in rarefactions.
The induced nematic order gives rise to the dielectric anisotropy 
\begin{equation}
	(\varepsilon_{zz}-\varepsilon_\perp)({\bf r},t) = - {\varepsilon_a\over 2}{\tilde G q^2\over A+(L+\tilde G)q^2}\delta\tilde\rho({\bf r},t).
    \label{anisotropy_sound}
\end{equation}
For a dynamic disturbance, $\tilde G$ is expected to increase substantially above its static value determined from the static fluctuations, as the inverse frequency becomes comparable to and falls below a characteristic dynamic time of the sources of the continuity equation Eq.~(\ref{tensorial}), where the disentanglement time \cite{grest1987,grest1989,sablic2016,sablic2017,sablic2017a} of the polymer chains seems to play an important role. Acoustically induced dielectric anisotropy Eq.~(\ref{anisotropy_sound}) is nevertheless small, as relative density variations due to acoustic excitations are normally tiny.

\begin{figure}[h]
  \centering
	\includegraphics[width=85mm]{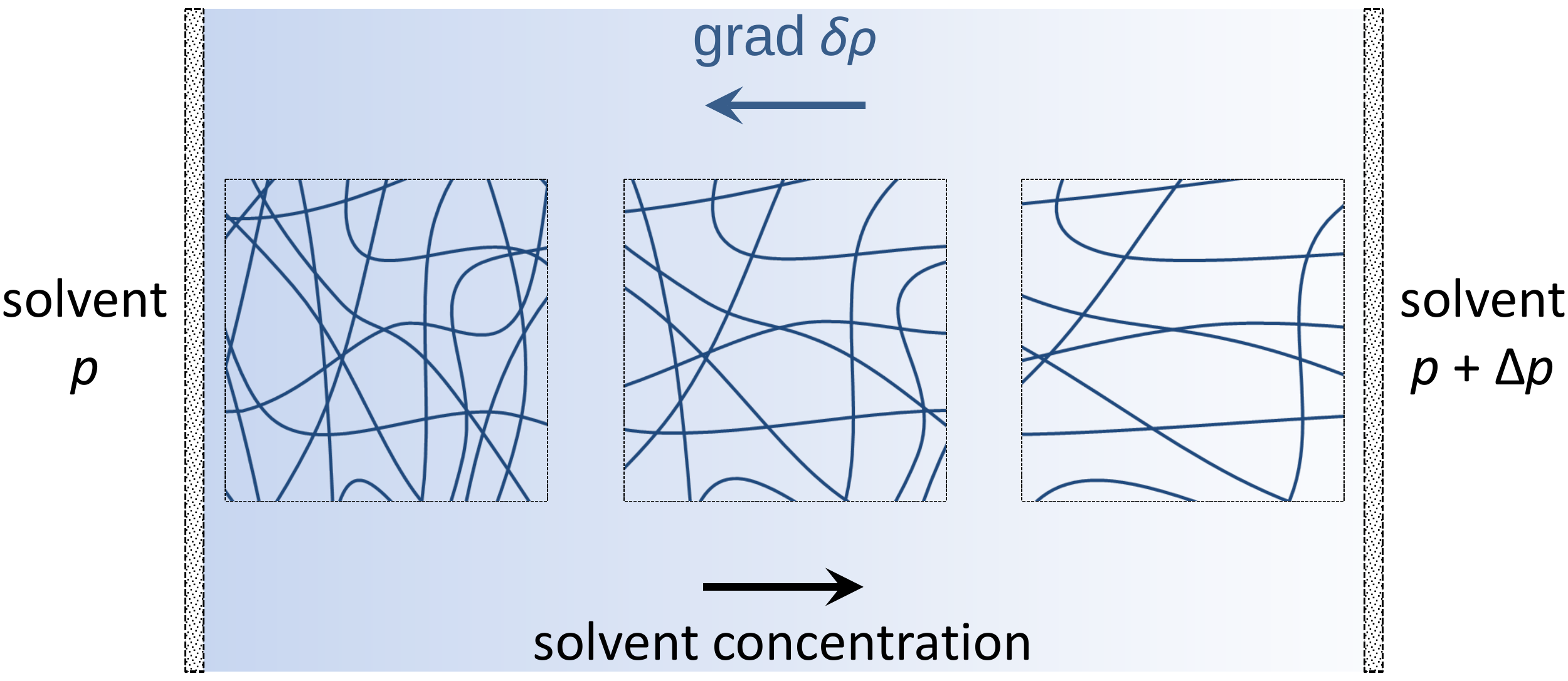} 
	\caption{Polymer concentration gradient $\nabla\delta\rho$ is schematically sustained by solvent pressure difference between a pair of semipermeable membranes.
If the chains are long and rather rigid, they accommodate the concentration gradient primarily by prolate/oblate orientational ordering.}
\label{fig:osmotic_stress}
\end{figure}

A much stronger effect can be expected in polymer solutions, where $\delta\rho$ represents polymer concentration variations rather than variations of the polymer melt density. 
Fig.~\ref{fig:osmotic_stress} presents a schematic situation, where a one-dimensional polymer concentration gradient $\nabla\delta\rho = \hat{\bf e}_z\partial_z\delta\rho$ is sustained by a difference in solvent osmotic pressure imposed by a pair of membranes at $z=\pm d/2$, permeable only to the solvent. 
Assuming that $(\partial_z\delta\rho)/\rho$ is small and constant, and choosing that the system be isotropic in the middle, $\delta Q_{zz}(z=0)=0$, we get an exponential profile of $\delta Q_{zz}$, Eq.~(\ref{sinh}), and hence of the induced dielectric anisotropy (see Appendix \ref{sec:birefringence} for details),
\begin{equation}
	(\varepsilon_{zz}-\varepsilon_\perp)(z) \approx - {\varepsilon_a\over 4} {G'\xi\rho\,\partial_z\delta\rho\over L+G'\rho^2}\,{\sinh (z/\xi)\over\cosh(d/(2\xi))},
\end{equation}
where $G' = G\left({\textstyle{2\over 3}} l_0\right)^2$ and $\xi\sim\sqrt{L/A}$ is the nematic correlation length.
That is, as a response to a constant concentration gradient, the nematic order and dielectric tensors are modulated in the boundary layers with characteristic thickness of the nematic correlation length $\xi$. Quite generally it is thus expected that, in the presence of a normal concentration gradient, any boundaries are typically {\it decorated} with short-range uniaxial nematic ordering/dielectric anisotropy in the normal direction. 

When the concentration gradient is not constant, however, nematic ordering is induced {\it globally}.
In lowest order, for the variation of the dielectric anisotropy, induced by $\partial_z\delta\rho(z)$ that varies slowly on the scale of $\xi$, we have from Eq.~(\ref{drive})
\begin{equation}
	\partial_z^2 (\varepsilon_{zz}-\varepsilon_\perp) \approx  -{\varepsilon_a\over 2}{G'\rho\over L+G'\rho^2}\partial^2_z\delta\rho.
    \label{epsilon_drive}
\end{equation}
A meaningful example is that with spherical symmetry: consider a spherical boundary of the solvent-rich side (the right boundary in Fig.~\ref{fig:osmotic_stress}), e.g., a membrane or an interface of a bubble, representing a spherical source of constant flux $\Phi$ of the solvent into the surrounding bulk (let the other boundary be absent or far away). Based on diffusion, it is reasonable to assume that for an overall constant density the profile of the polymer concentration at a sufficient distance $r$ from the center of the bubble will be 
\begin{equation}
	\rho(r) = \rho_0-\Phi/ (4\pi D r),
    \label{rho_bubble}
\end{equation}
with $\rho_0$ its bulk concentration and $D$ the diffusivity.
Consequently, for a system which is isotropic in the bulk we expect that such spherical inhomogeneity will induce uniaxial ordering of the chains in the radial direction, Eq.~(\ref{masterpiece_solution}), as derived in lowest order in Appendix \ref{sec:birefringence}. The corresponding dielectric anisotropy in the limit of sufficiently large $r$ is then
\begin{equation}
	(\varepsilon_{rr}-\varepsilon_\perp)(r)\approx -\varepsilon_a{G'\rho_0\over 6L + G'\rho_0^2}\,
    {\Phi\over 4\pi D r}.
    \label{epsilon_bubble}
\end{equation}
Interestingly, in the considered regime the induced nematic ordering Eq.~(\ref{masterpiece_solution}) is negative, i.e., oblate, rather than prolate as one might have naively guessed from the decreasing polymer concentration.
Note that in all of the expressions Eqs.~(\ref{anisotropy_sound})-(\ref{epsilon_bubble}) the polymer density appears in the form of the product $\rho_{(0)} l_0$, i.e., the total length of the polymer per unit volume, which is the relevant volume density for the coupling.

Such optical effects due to the dielectric anisotropy, resulting from concentration variations in solutions of long linear polymers with limited flexibility, should be observable, e.g., as an osmotic birefringence. The phenomenon is akin to shear flow-induced birefringence in fluid polymers and stress-optic law in elastic solid dielectric materials, i.e, the direct coupling between the strain and the dielectric tensors. The key distinction is, however, that the osmotic-stress-induced birefringence takes place in a static liquid, where there exists no strain/strain rate tensor that could couple to the dielectric tensor. In this case, the coupling --- which is a manifestation of the microscopic geometrical constraint Eq.~(\ref{mic}) --- is via the concentration gradient through the tensorial continuity equation Eq.~(\ref{tensorial}).

It is challenging to determine the macroscopic parameters in Eq.~(\ref{f}) of real polymeric systems by conducting microscopic simulations with known force fields, and fitting the extracted fluctuation amplitudes with theoretical expressions as we have done for our model system. In this respect, DNA is of particular interest: a recently developed open-boundary molecular dynamics of a DNA molecule in explicit salt--hybrid explicit/implicit water solution \cite{zavadlav2018} would enable DNA simulations at lower, physiological salt concentrations $\sim 0.15$\,M.
At this stage, let us make an estimate of the coupling strength for DNA, making use of the theoretical prediction Eq.~(\ref{G_xip}).
Assuming a persistence length $\xi_p \sim 50$\,nm and chains much longer than that, a sub-nematic DNA concentration 5\,mg/ml (with the atomic mass of 650\,dalton/base pair and the length 0.33\,nm/base pair this amounts to a total length per unit volume of $\rho_0 l_0\approx 1.5$\,mm/$\mu$m$^3$) and temperature 300\,K, one gets $\tilde G = G({2\over 3} \rho_0 l_0)^2 \approx {2\over 3}\kT\xi_p \rho_0 l_0 \approx 2\times 10^{-13}$\,N. This is expected to be at least comparable if not much larger than the elastic constant $L$ (in the case of the simulated $N_{\rm s}=128$ chains, we get $\tilde G/L \sim 20$). Consequently, the expected induced nematic ordering, Eq.~(\ref{drive}), is of the same order of magnitude as the relative variation of the polymer concentration.

\section{Conclusions}

\noindent
We have established a tensorial description of coupled density and nematic order fluctuations in the isotropic phase of linear polymer melts/solutions. We have validated and confirmed the devised tensorial conservation law, which connects density and orientational order, by conducting extensive Monte Carlo simulations of isotropic polymer melts composed of worm-like chains with variable length and stiffness. 
As demonstrated, due to this coupling orientational order is induced by density/concentration variations, which can be macroscopically relevant even in the otherwise isotropic polymeric liquid.
Our results show that these effects become increasingly important in particular as the chains get stiffer.
Rather surprisingly, the coupling is notable also for extremely short chains (a few monomers).
Moreover, we have alluded to the possible relevance of symmetry-allowed coupling between chain curvature and other system variables, and presented means of its quantification. 

The average chain curvature as a macroscopic variable is not normally encountered. We have presented a phenomenon where it is relevant for the static macroscopic response of line polymers. As such, it is accessible from the macroscopic level and can be in principle measured experimentally, if not determined directly by microscopic simulations.
Our multiscale formalism is general and robust and provides a way to construct bottom-up continuum models of polymer melts/solutions, e.g., dense DNA columnar phases, by allowing us to extract the unknown parameters of the continuum models from underlying microscopic simulations.

\appendix

\section{Tensorial conservation law}
\label{sec:tensorial}

\noindent
A linear polymer is modeled as a continuous WLC presented by the microscopic density field of the polymer length
\begin{equation}
	\rho^{\rm mic}({\bf x}) =  \sum_\alpha\int_{{\bf x}^\alpha(s)} {\rm d}s ~\delta({\bf x} - {\bf x}^\alpha(s)),
    \label{rhomic}
\end{equation}
i.e., the total length of the polymer per unit volume,
where ${\bf x}^\alpha(s)$ is the contour of the chain $\alpha$ in natural parametrization. For brevity we will be omitting the superscript $^\alpha$ and the sum $\sum_\alpha$ over the chains. The continuity of ${\bf x}(s)$ stands for the unbroken connectivity of the polymer chains.
A microscopic traceless polymer nematic tensor field can be defined as
\begin{eqnarray}
	J_{ij}^{\rm mic}({\bf x}) &=&  \int_{{\bf x}(s)} {\rm d}s ~\delta({\bf x} - {\bf x}(s))\, 
				{\textstyle{3\over 2}}\left[t_i(s) t_j(s)-{\textstyle{1\over 3}}\delta_{ij}\right]\nonumber\\ 
                &\equiv&
				\tilde{J}^{\rm mic}_{ij}({\bf x}) - \textstyle{1\over 2}\delta_{ij}\rho^{\rm mic}({\bf x}),
	\label{J_traceless_mic}
\end{eqnarray}
where ${\bf t}(s) = {{\rm d}{\bf x}(s)/{\rm d}s}$ is the unit tangent on the chain.

Taking a divergence of Eq.~(\ref{J_traceless_mic}),
\begin{eqnarray}
	&&\partial_j\left({J}_{ij}^{\rm mic} + {\textstyle{1\over 2}}\delta_{ij}\rho^{\rm mic}\right) =\\ &&\quad\qquad{\textstyle{3\over 2}}
	\int_{{\bf x}(s)} {\rm d}s ~{{\rm d}x_i(s)\over {\rm d}s}{{\rm d}x_j(s)\over {\rm d}s} {\partial\over\partial x_j}\delta({\bf x} - {\bf x}(s)),\nonumber
\end{eqnarray}
using ${{\rm d}x_j(s)\over {\rm d}s}{\partial\over\partial x_j}\delta({\bf x} - {\bf x}(s)) = -{{\rm d}x_j(s)\over {\rm d}s}{\partial\over\partial x_j(s)}\delta({\bf x} - {\bf x}(s)) = -{{\rm d}\over{\rm d}s}\delta({\bf x} - {\bf x}(s))$ and integrating by parts, we get
\begin{eqnarray}
	&&\partial_j\left({J}_{ij}^{\rm mic} + {\textstyle{1\over 2}}\delta_{ij}\rho^{\rm mic}\right) =    \label{mic} \\ 
    &&\qquad\qquad{\textstyle{3\over 2}}\left[t_i(0)\delta({\bf x}-{\bf x}(0))-t_i(L)\delta({\bf x}-{\bf x}(L))\right]\nonumber\\  
    &&\qquad\qquad+ {\textstyle{3\over 2}} \int_{{\bf x}(s)} {\rm d}s ~{{\rm d}^2x_i(s)\over {\rm d}s^2}\delta({\bf x} - {\bf x}(s)),\nonumber
\end{eqnarray}
where $s=0$ and $s=L$ corresponds to the beginning and ending of a chain, respectively. In the absence of polar order, this identification is arbitrary and the beginning and ending tangents can be unified into a single type of tangents ${\bf t}^{\rm n}$ always pointing inwards, such that
\begin{eqnarray*}
	&&t_i(0)\delta({\bf x}-{\bf x}(0))-t_i(L)\delta({\bf x}-{\bf x}(L)) =\\ 
 &&t_i^{\rm n}(0)\delta({\bf x}-{\bf x}(0))+t_i^{\rm n}(L)\delta({\bf x}-{\bf x}(L)).
\end{eqnarray*}

Writing a microscopic field ${\bf F}^{\mathrm{mic}}({\bf x})$, i.e.,\ Eq.~(\ref{rhomic}), Eq.~(\ref{J_traceless_mic}) and also the last term of  Eq.~(\ref{mic}), in the form
\begin{equation}
	{\bf F}^{\mathrm{mic}}({\bf x}) = \int_{{\bf x}(s)} {\rm d}s ~\delta({\bf x} - {\bf x}(s)) ~{\bf f}({\bf x}(s)),
\end{equation}
coarse-graining it to the mesoscopic volume $V$ centered at $\bf x$ (denoted by $\overbracket{\phantom{F^{\mathrm{mic}}}}$~) gives the corresponding mesoscopic field \cite{svensek2013}
\begin{eqnarray}
	&&{\bf F}({\bf x}) = \overbracket{{\bf F}^{\mathrm{mic}}}({\bf x})=
			   {1\over V}\int_{V({\bf x})} {\rm d}^3 x' ~{\bf F}^{\mathrm{mic}}({\bf x}') =\\
			   &&{1\over V}\!\int_{{\bf x}(s)\in V({\bf x})}\!\!\!\! {\rm d}s ~{\bf f}({\bf x}(s)) =
			   {L({\bf x})\over V} {1\over L({\bf x})}\!\int_{{\bf x}(s)\in V({\bf x})}\!\!\!\! {\rm d}s ~{\bf f}({\bf x}(s)), \nonumber
\end{eqnarray}
where $L({\bf x})=\int_{{\bf x}(s)\in V({\bf x})}~{\rm d}s \equiv N({\bf x}) l_0$ is the total length of the chain within the volume $V$, which can be expressed in terms of an arbitrary segment length $l_0$ and the number $N$ of these segments within the volume. 
Hence, the mesoscopic field can be written as
\begin{equation}
	{\bf F}({\bf x}) = \rho({\bf x}) l_0~ \bar{{\bf f}}({\bf x}),
	\label{F_meso}
\end{equation}
where $\rho({\bf x}) = {N({\bf x})/ V}$ is the mesoscopic volume number density of the segments and 
\begin{equation}
	\bar{{\bf f}}({\bf x}) = {1\over L({\bf x})}\int_{{\bf x}(s)\in V({\bf x})} {\rm d}s ~{\bf f}({\bf x}(s))
\end{equation}
is the mesoscopic average of ${\bf f}({\bf x}(s))$.

Applying this coarse-graining procedure to Eq.~(\ref{mic}), where in particular $\overbracket{\rho^{\rm mic}}=\rho l_0$, $\overbracket{J_{ij}^{\rm mic}}\equiv J_{ij}=\rho l_0 Q_{ij}$, and bearing in mind that the coarse-graining and $\nabla$ commute,
we get an equation for continuum mesoscopic fields --- the tensorial conservation law
\begin{equation}
	\partial_j\left[\rho(Q_{ij}+{\textstyle{1\over 2}}\delta_{ij})\right] = {\textstyle{3\over 2}{1\over l_0}}g_i + {\textstyle{3\over 2}}\rho k_i,
\end{equation}
where $\sf Q$ is the nematic order tensor, 
\begin{equation}
	Q_{ij}({\bf x}) = {1\over L({\bf x})}\int_{{\bf x}(s)\in V({\bf x})} {\rm d}s ~{\textstyle{3\over 2}}\left[t_i(s) t_j(s)-{\textstyle{1\over 3}}\delta_{ij}\right],
\end{equation}
${\bf g}({\bf x}) = \overbracket{{\bf t}^{\rm n}(0)\delta({\bf x}-{\bf x}(0))+{\bf t}^{\rm n}(L)\delta({\bf x}-{\bf x}(L))}$ is the mesoscopic density of chain-end tangents
and 
\begin{equation}
	{\bf k}({\bf x}) = {1\over L({\bf x})}\int_{{\bf x}(s)\in V({\bf x})} {\rm d}s~ {{\rm d}^2{\bf x}(s)\over{\rm d}s^2}
    \label{curvature}
\end{equation}
is the mesoscopic average chain curvature vector.

\section{Free-energy diagonalization and equipartition}
\label{sec:diagonalization}

\noindent
The real expressions $\lambda^\pm$ and $a_\pm$ in Eq.~(\ref{fqdiag}) read
\begin{eqnarray}
	a_\pm &=& \frac{1}{4 \tilde G q^2}\Bigg\{4\tilde B-12 A-q^2 (7 \tilde G+12 L) \label{apm}\\ 
    &\pm& \sqrt{\left[4\tilde B-12 A-q^2 (7 \tilde G+12 L)\right]^2 + 32\tilde G q^4}\Bigg\},\nonumber
\end{eqnarray}
\begin{eqnarray}
	\lambda^\pm &=& \frac{1}{16} \Bigg\{4 \tilde B+12 A+3 q^2 (3 \tilde G+4 L) \label{lampm}\\
    &\pm&\sqrt{\left[4\tilde B-12 A-q^2 (7 \tilde G+12 L)\right]^2 + 32\tilde G q^4}\Bigg\},\nonumber
\end{eqnarray}
where $\tilde B = B+B'q^2$.
The stability condition for the free energy Eq.~(\ref{fqdiag}) requires $\lambda^\pm>0$.

The last two terms in Eq.~(\ref{fqdiag}) represent the contributions of two coupled fluctuation modes. 
Alternatively, one can write them as
\begin{equation}
	{\lambda^+\lambda^-(a_+-a_-)^2\over\lambda^+ v_-^2+\lambda^- v_+^2}|\delta\tilde\rho|^2 + |\dots\delta Q_{zz}+\dots\delta\tilde\rho|^2
\end{equation}
or
\begin{equation}
	{\lambda^+\lambda^-(a_+-a_-)^2\over\lambda^+ v_-^2 a_+^2+\lambda^- v_+^2 a_-^2}|\delta Q_{zz}|^2 + |\dots\delta\tilde\rho\dots\delta Q_{zz}|^2
\end{equation}
and therefrom calculate $\langle|\delta Q_{zz}|^2\rangle$ and $\langle|\delta\tilde\rho|^2\rangle$, Eqs.~(\ref{Qzzfluct}) and (\ref{rhofluct}).

Using Eqs.~(\ref{Qzzfluct}) and (\ref{rhofluct}) together with the average of one of the last terms of Eq.~(\ref{fqdiag}), e.g.,
\begin{equation}
	\left\langle \left|a_\pm\delta\tilde\rho+\delta Q_{zz}\right|^2 \right\rangle = {\kT\over 2}V{v_\pm^2\over\lambda^\pm}
\end{equation}
which we are not giving explicitly, one finally arrives at the cross-correlation Eq.~(\ref{cross-correlation}).

\section{Mesoscopic WLC model and numerical simulation details}
\label{sec:simulation}

\noindent
In the MC simulations, we use a recently developed mesoscopic model of discrete WLCs \cite{gemunden2015,popadic2018}.
The modeled system contains $N_{\rm c}$ WLCs comprised of $N_{\rm s}$ linearly connected segments of fixed length $l_0$. Consecutive
segments are subjected to a standard angular potential
\begin{equation}
	U_{\rm b} = -\epsilon {\bf u}^{i,s}\cdot {\bf u}^{i,s+1},
    \label{angular_potential}
\end{equation}
where 
${\bf u}^{i,s}$
is the unit vector along the $s$-th segment of the $i$-th chain and
$\epsilon$ controls the WLC bending stiffness. Non-bonded interactions between segments are introduced via
the potential $U_{\rm nb} = \kappa U(r_{ij}^{st})
$, where $\kappa$ is the strength of the isotropic repulsion between the segments and 
$U(r_{ij}^{st}) = C_{0}\Theta \left(2\sigma - r_{ij}^{st}\right)\left[4\sigma+r_{ij}^{st}\right]\left[2\sigma-r_{ij}^{st}\right]^2$ represents the overlap of two spherical clouds centered on the $s$-th and $t$-th segments of the $i$-th and $j$-th chain, respectively;
$r_{ij}^{st}$ is the distance between the segments and $\sigma$ controlls the interaction range as indicated by the Heaviside function $\Theta$. To verify the predictions of the macroscopic theory
it is sufficient to employ a generic model with a single ``microscopic'' length scale. Hence, we set $\sigma = l_0$, although other choices are
possible~\cite{daoulas2012,greco2016} when modeling actual materials.
The normalization constant of $U(r_{ij}^{st})$ is set to $C_{0} = 3l_0^3/(64\pi \sigma^6)$. 
We empirically set $\kappa = 7.58\,k_{\rm B} T$ \cite{gemunden2015}.
Several molecular flexibilities ranging from $\epsilon = 0$ to $\epsilon=13.136\,k_{\rm B} T$ are addressed, corresponding to decreasing flexibility of the chains.
The MC algorithm utilizes a combination of standard random monomer displacement and slithering snake moves \cite{frenkel2001}.
In addition, every $N_0=N_{\rm c} N_{\rm s}$ random displacement and slithering snake moves, a volume fluctuation move at pressure $Pl_0^3/(\kT) = 2.87$ is employed \cite{tuckerman2010}.

The fluctuations of any variables $\delta a({\bf q})=\sum_s a_s {\rm e}^{-{\rm i}{\bf q}\cdot{\bf r}^s}$ and $\delta b({\bf q})=\sum_s b_s {\rm e}^{-{\rm i}{\bf q}\cdot{\bf r}^s}$ are extracted via their correlation functions,
\begin{eqnarray}
	&&{1\over N_0}{1\over 2}\big[\langle \delta a({\bf q})\delta b(-{\bf q})\rangle + \langle\delta a(-{\bf q})\delta b({\bf q})\rangle\big] = \label{corrsim}\\
	&&\quad{1\over N_0}\Bigg\langle \bigg[\sum_s a_s\cos({\bf q}\cdot{\bf r}^s)\bigg] 
									 \bigg[\sum_s b_s\cos({\bf q}\cdot{\bf r}^s)\bigg] + \nonumber\\
									 &&\qquad\quad~\bigg[\sum_s a_s\sin({\bf q}\cdot{\bf r}^s)\bigg] 
									 \bigg[\sum_s b_s\sin({\bf q}\cdot{\bf r}^s)\bigg]
					\Bigg\rangle,\nonumber
\end{eqnarray}
where $s=1 \dots N_0$ runs over the segments of all chains and ${\bf r}^s$ are their positions.
For segment density fluctuations $\delta\rho$ we have $a_s = 1$, and for the nematic fluctuations $\delta J_{ij}$ we have $a_s = (3u_i^s u_j^s-\delta_{ij})/2$.
Note that the coarse graining does not affect the ${\bf q}\to 0$ Fourier components, or in other words, the ${\bf q}\to 0$ components of the extracted discrete variables are automatically coarse-grained. Hence, the long wavelength correlations Eq.~(\ref{corrsim}) computed from the simulation data can be directly compared to the predictions of the continuum theory Eqs.~(\ref{Qxyfluct})-(\ref{cross-correlation}).

The ensemble volume is free to fluctuate and the set of $\bf q$ vectors is determined by the current box size.
Since the system is isotropic, all quantities depend only on the magnitude $|{\bf q}|=q$. We average them over spherical shells with thickness $\Delta q\sim 2\pi/\langle L\rangle$, taking care that also the smallest shells ($q\to 0$) are adequately populated.
In an isotropic system, the isotropic symmetry of non-scalar quantities is broken only by the direction $\bf q$, which is exploited in the averaging procedure as follows. For every $\bf q$, we set the coordinate system such that ${\bf q} = q\hat{\bf e}_z$ as we have done in Eqs.~(\ref{Qxyfluct})-(\ref{cross-correlation}), while  
$$
\hat{\bf e}_x = {\hat{\bf e}_{x'}-(\hat{\bf e}_{x'}\cdot\hat{\bf e}_{z})\hat{\bf e}_{z}\over
	  |\hat{\bf e}_{x'}-(\hat{\bf e}_{x'}\cdot\hat{\bf e}_{z})\hat{\bf e}_{z}|} 
$$      
and $\hat{\bf e}_y = \hat{\bf e}_z\times\hat{\bf e}_x$,  
where $\hat{\bf e}_{x'}$ is aligned with the simulation box. With that, for the component $\delta J_{zz}$ we have $a_s = [3({\bf u}^s\cdot\hat{\bf e}_z)^2-1]/2$ and for the components $\delta J_{\{x,y\}z}$ we have $a_s = 3({\bf u}^s\cdot\hat{\bf e}_{\{x,y\}})({\bf u}^s\cdot\hat{\bf e}_z)/2$. 

The computed correlations are then averaged over collected configurations.
When calculating averages, we use block-averaging with block size $\tau$, where $\tau$ is the number of MC steps needed to decorrelate the end-to-end vector of the WLC \cite{allen1990}.

\section{Bend of monomer pairs}
\label{sec:bend}

\noindent
Here, we write down the general statistical result for the bending configuration of an independent monomer pair with bending energy Eq.~(\ref{F(theta)}). 

In the stiff limit ($\kT/\epsilon\ll 1$), the result Eq.~(\ref{k01}) holds for the individual component of the curvature vector perpendicular to the monomers, and Eq.~(\ref{<k2>}) holds for the curvature in any direction in the case of isotropically averaged monomer orientation.

Conversely, in the ideally flexible limit ($\kT/\epsilon\gg 1$) the chain undergoes a random walk. 
Putting ${\bf u}^i=\hat{\bf e}_z$ and ${\bf u}^{i+1}= {\bf u}^i + l_0 {\bf k}_0 = \hat{\bf e}_r$, we have 
\begin{equation}
	(l_0 k_0)^2 = 2(1-\cos\theta)
    \label{k0_finite}
\end{equation}
and the solid angle average over all possible orientations of ${\bf u}^{i+1}$ is $l_0^2\overline{k_0^2} = 2$.
Hence, in the isotropically averaged situation, the average square of the curvature in any direction is 
\begin{equation}
	\langle k_1^2\rangle^{\infty} = {1\over 3}\overline{k_0^2}={2\over3}{1\over l_0^2},
    \label{<k2>_random}
\end{equation}
where the superscript $^\infty$ stands for $\kT/\epsilon\to\infty$.

\begin{figure}[h]
  \centering
    \includegraphics[]{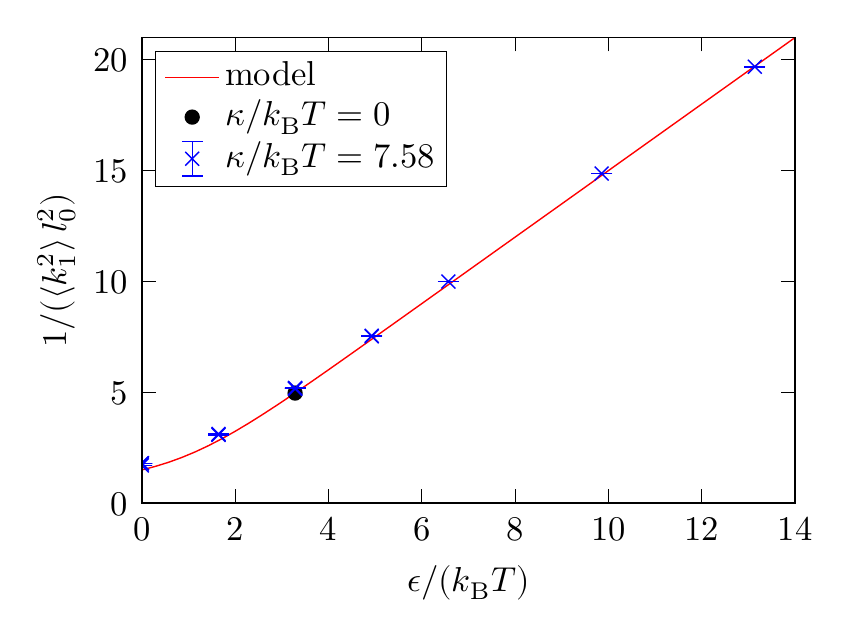}
    \caption{The values of $1/\langle k_1^2\rangle$ from the simulations (points) for all studied chain lengths $N_{\mathrm{s}}$, and plot (no fitting) of the model Eq.~(\ref{effective_stiffness}). The extra circular point belongs to the simulation without the repulsive potential ($\kappa=0$) and falls exactly onto the theoretical curve.
    }
\label{fig:1/k12}
\end{figure}

For general flexibility,
the partition function corresponding to the energy Eq.~(\ref{F(theta)}) is
\begin{equation}
	Z = \int_{-1}^1\! {\rm d}(\cos\theta)\,{\rm e}^{-\beta\Delta F(\theta)} = {\rm e}^{-\beta\epsilon}{2\over\beta\epsilon}\sinh\beta\epsilon,
\end{equation}
where $\beta=1/(\kT)$, and the average energy is
\begin{equation}
	\langle\Delta F\rangle = -{\partial\over\partial\beta}\ln Z = {1\over\beta}-\epsilon\left(\coth\beta\epsilon-1\right).
\end{equation}
With that, using Eqs.~(\ref{F(theta)}) and (\ref{k0_finite}) and furthermore taking into account the isotropy as done in Eq.~(\ref{<k2>_random}), the average square of the curvature in an arbitrary direction is
$\langle k_1^2\rangle = 2/(3\epsilon l_0^2)\,\langle\Delta F\rangle$, such that
\begin{equation} 
   {1\over\langle k_1^2\rangle} = {3\over 2} l_0^2 {\epsilon\over \kT - \epsilon\left({\rm \coth}{\epsilon\over \kT}-1\right)}.
   \label{effective_stiffness}
\end{equation}
One can verify that the result Eq.~(\ref{effective_stiffness}) includes both the  stiff chain limit Eq.~(\ref{<k2>}) and the flexible chain limit Eq.~(\ref{<k2>_random}).
As it turns out (see Fig.~\ref{fig:1/k12}), this exact statistical result for the isolated segment pair applies with great accuracy also to pairs surrounded by neighbouring chains of the simulated melt.

Finally, the quantity $\langle k_1^2\rangle$ can be determined from simulation data by measuring the average square of the curvature in an arbitrarily chosen direction. Fig.~\ref{fig:1/k12} reveals that the agreement between Eq.~(\ref{effective_stiffness}) and $1/\langle k_1^2\rangle$ from the simulations is striking. Moreover, the additional point in Fig.~\ref{fig:1/k12} with the repulsive potential between all monomers switched off ($\kappa=0$) indicates that the repulsion from other chains as well as the repulsion between the monomers forming the joint (correction to the bending rigidity $\epsilon$) is small. Also small is apparently the influence of the continuity equation Eq.~(\ref{tensorial_fluct}) on the magnitude $|\delta {\bf k}|^2$ of its curvature source. Were this not the case, the large compressibility when $\kappa=0$ would result in a significant change of $\langle k_1^2\rangle$, as a result of the vanishing cost of density fluctuations that are coupled with fluctuations $\delta{\bf k}$ through the constraint Eq.~(\ref{tensorial_fluct}).

\section{Induced orientational order}
\label{sec:birefringence}

\noindent
The equilibrium coupling of the nematic order tensor to a given fixed density or concentration variation $\delta\rho({\bf r})$ is obtained by minimizing the part of the free energy Eq.~({\ref{f}}) belonging to $\delta Q_{ij}$ and the sources of the continuity constraint Eq.~(\ref{tensorial}),
\begin{eqnarray}
	f' &=& {1\over 2} A \left(\delta Q_{ij}\right)^2 + {1\over 2} L \left(\partial_k\delta Q_{ij}\right)^2 \label{f'}\\
    &+& {1\over 2}G'\left[\rho\partial_j\delta Q_{ij}+(\delta Q_{ij}+{\textstyle{1\over 2}}\delta_{ij})\partial_j\delta\rho\right]^2, \nonumber
\end{eqnarray}
where $G' = G\left({\textstyle{2\over 3}} l_0\right)^2$, with the result
\begin{eqnarray}
	{\delta f'\over\delta(\delta Q_{ij})} = 0 &=& A\,\delta Q_{ij} - L\,\partial_k^2\delta Q_{ij} \label{extreme}\\
    &-&G'\rho\, \partial_j\!\left[\rho\partial_k\delta Q_{ik}+(\delta Q_{ik}+{\textstyle{1\over 2}}\delta_{ik})\partial_k\delta\rho\right],\nonumber
\end{eqnarray}
\begin{eqnarray}
    &&p_k {\partial f'\over\partial(\partial_k\delta Q_{ij})}\Big\vert_\partial = 0 =\label{extreme_boundary}\\ 
    &&L\,\partial_k\delta Q_{ij}\big\vert_\partial\,p_k 
    +G'\rho\left[\rho\partial_k\delta Q_{ik}+(\delta Q_{ik}+{\textstyle{1\over 2}}\delta_{ik})\partial_k\delta\rho\right]\!\big\vert_\partial\, p_j,\nonumber
\end{eqnarray}
where, in case relevant, Eq.~(\ref{extreme_boundary}) holds at the bounding surface with the normal $\bf p$. For the reason of generality, the non-linearized continuity constraint Eq.~(\ref{tensorial}) has been considered in Eq.~(\ref{f'}).

We have already seen via the Fourier space that a density (acoustic) plane wave $\delta\tilde\rho({\bf r},t)=\delta\rho({\bf r},t)/\rho_0$ with wave vector ${\bf q}=q\hat{\bf e}_z$ couples only to the component $\delta Q_{zz}$. 
Taking $\nabla=\hat{\bf e}_z\partial_z$ in  Eq.~(\ref{extreme}) and linearizing it, we get
\begin{equation}
	\delta Q_{zz}({\bf r},t) = - {1\over 2}{\tilde G q^2\over A+(L+\tilde G)q^2}\delta\tilde\rho({\bf r},t).
    \label{Qzz_sound}
\end{equation}

In polymer solutions, $\delta\tilde\rho$ represents concentration variations rather than variations of the density. In the one-dimensional case where the externally imposed concentration gradient is along $z$, Eq.~(\ref{extreme}) leads to
\begin{eqnarray}
	(L+G'\rho^2)\partial_z^2\delta Q_{zz}+2G'\rho(\partial_z\delta\rho)\partial_z\delta Q_{zz} + && \label{laplace_rho} \\
    G'(\delta Q_{zz}+{\textstyle{1\over 2}})\rho\partial_z^2\delta\rho - A\delta Q_{zz} &=& 0.\nonumber
\end{eqnarray}
If the relevant range is $-d/2<z<d/2$ and we for simplicity assume that $|\partial_z\delta\rho| d/\rho\ll 1$, the homogeneous solution of Eq.~(\ref{laplace_rho}), i.e., the solution for constant $\partial_z\delta\rho$, is of the simple form $\delta Q_{zz}(z) = C_1{\rm e}^{\lambda_1 z} + C_2{\rm e}^{\lambda_2 z}$, with real
\begin{equation}
	\lambda_{1,2} = {-G'\rho\partial_z\delta\rho \pm \sqrt{(G'\rho\partial_z\delta\rho)^2+A(L+G'\rho^2)}\over L+G'\rho^2}.
\end{equation}
The relevant regime is that of weak density--nematic coupling and small concentration gradient, such that $(G'\rho\partial_z\delta\rho)^2\ll{A(L+G'\rho^2)}$ holds and $\lambda_{1,2}\to \pm \sqrt{A/L}=\pm\xi^{-1}$. Note that this limit is equivalent to linearizing Eqs.~(\ref{extreme})-(\ref{extreme_boundary}) with respect to $\delta Q_{ij}$ and $\delta\rho$.
Hence, in a good approximation the solution is further simplified, and with the choice $\delta Q_{zz}(z=0)=0$ becomes
\begin{equation}
	\delta Q_{zz}(z) \approx C_1\sinh\lambda_1 z.
\end{equation}
The boundary condition at $z=d/2$ (or, equivalently, $z=-d/2$) follows from linearized Eq.~(\ref{extreme_boundary}), 
\begin{equation}
	\partial_z\delta Q_{zz}(d/2) = - {1\over 4}{G'\rho\partial_z\delta\rho\over L+G'\rho^2},
\end{equation}
so that finally we have
\begin{equation}
	\delta Q_{zz}(z) \approx -{G'\rho\,\partial_z\delta\rho\over 4\lambda_1(L+G'\rho^2)}\,{\sinh\lambda_1 z\over\cosh(\lambda_1 d/2)}.
    \label{sinh}
\end{equation}
That is, as a response to a constant concentration gradient, $\delta Q_{zz}$ is modulated in the boundary layers with characteristic thickness of the nematic correlation length $\xi$.

When the concentration gradient is not constant, $\partial_z^2\delta\rho$ presents an inhomogeneity in Eq.~(\ref{laplace_rho}). If $\partial_z\delta\rho(z)$ is a slowly varying function (on the scale of $\xi$), we get from Eq.~(\ref{laplace_rho}) in the limit $\delta Q_{zz}\to 0$ 
\begin{equation}
	\partial^2_z Q_{zz} \approx -{1\over 2}{G'\rho\over L+G'\rho^2}\partial^2_z\delta\rho,
    \label{drive}
\end{equation}
which thus represents the drive of the induced nematic ordering.

If the polymer concentration is spherically symmetric, one expects a uniaxial ordering of the chains in the radial direction and can write, without loss of generality, in spherical coordinates $(r,\theta,\phi)$
\begin{eqnarray}
	{\sf Q}({\bf r}) &=& Q_{rr}(r)\left[\hat{\bf e}_r\otimes\hat{\bf e}_r-{\textstyle{1\over 2}}\left(\hat{\bf e}_\theta\otimes\hat{\bf e}_\theta+\hat{\bf e}_\phi\otimes\hat{\bf e}_\phi\right)\right]\nonumber\\ 
    &\equiv& Q_{rr}(r)\,{\sf T}({\bf r}). 
\end{eqnarray}
With $\nabla = \hat{\bf e}_r{\partial\over\partial r} + \hat{\bf e}_\theta{\partial\over r\partial\theta} + \hat{\bf e}_\phi{\partial\over r\sin\theta\partial\phi}$, the nonzero derivatives of the spherical base vectors and some algebra we find the auxiliary expressions
\begin{eqnarray}
	\nabla\cdot{\sf Q} &=& 
	\left(\partial_r Q_{rr}+{3\over r}Q_{rr}\right)\hat{\bf e}_r,\\
    \nabla^2 {\sf Q} &=& 
    {1\over r^2}{\partial\over\partial r}\left(r^2{\partial Q_{rr}\over\partial r}\right){\sf T} - {6\over r^2}Q_{rr}{\sf T}.
\end{eqnarray}
It is sufficient to take only the $\hat{\bf e}_r$-part of the gradient (denoted $\partial_j$) in Eq.~(\ref{extreme}),  
\begin{eqnarray}
	&&G'\rho\, \partial_r\left(\rho\partial_r Q_{rr}+{3\over r}\rho Q_{rr}+(Q_{rr}+{\textstyle{1\over 2}})\partial_r\rho\right)\hat{\bf e}_r\otimes \hat{\bf e}_r  \nonumber\\ 
    &&~ -A Q_{rr}{\sf T} +{1\over r^2} L\left[\partial_r(r^2\partial_r Q_{rr})-6Q_{rr}\right]{\sf T} = 0,
    \label{masterpiece}
\end{eqnarray}
and consider the component along $\hat{\bf e}_r\otimes\hat{\bf e}_r$, which one gets simply by dropping all tensors in Eq.~(\ref{masterpiece}), since $\hat{\bf e}_r\cdot{\sf T}\cdot\hat{\bf e}_r=1$.

If the polymer concentration is of the specific form given by Eq.~(\ref{rho_bubble}), we have
$\partial^2_r\rho = -2\Phi/(4\pi D r^3)$. 
Using the same approximations as for Eq.~(\ref{drive}), which are $\rho\approx\rho_0$, $\left\{(G'\rho_0\partial_r\rho)^2, (G'\rho_0^2/r)^2\right\} \ll{A(L+G'\rho^2)}$ and $Q_{rr}\equiv\delta Q_{rr}\to 0$, from Eq.~(\ref{masterpiece}) we finally get 
\begin{eqnarray}
	&&(L+G'\rho_0^2)\,\partial_r^2\delta Q_{rr} + {2L+3G'\rho_0^2\over r}\partial_r \delta Q_{rr}-{6L\over r^2}\delta Q_{rr} \nonumber\\
    &&\qquad\qquad\qquad\qquad\qquad\qquad\quad = G'\rho_0{\Phi\over 4\pi D r^3}.
\end{eqnarray}
Using a power-law ansatz, the solution is
\begin{equation}
	\delta Q_{rr}(r)\approx -{G'\rho_0\over 6L + G'\rho_0^2}\,
    {\Phi\over 4\pi D r}.
    \label{masterpiece_solution}
\end{equation}

\section*{Acknowledgments}
\noindent
Our special appreciation and gratitude goes to Kostas Ch.~Daoulas for contributing the soft WLC model, sharing the simulation code, offering his experience, and participating in numerous valuable discussions throughout this long-lasting investigation. 
We acknowledge financial support through grants P1-0002 and J1-7435 from the Slovenian Research Agency.

\bibliography{literature}

\end{document}